\documentclass[11pt,a4paper]{article}
\usepackage{graphicx} 
\usepackage{tabularx} 
\usepackage{array}
\usepackage{adjustbox}
\usepackage[table]{xcolor}
\newcommand{\uparrowgreen}{\textcolor{green!60!black}{$\uparrow$}}
\newcommand{\downarrowred}{\textcolor{red!70!black}{$\downarrow$}}

\usepackage{tikz}

\newcommand{\tableuparrowgreen}{%
  \hspace{0.4em}%
  \tikz[baseline=0pt, x=1ex, y=1.7ex]{%
    \fill[green!60!black] (-0.42,0.55) -- (0,1.00) -- (0.42,0.55) -- cycle;
    \fill[green!60!black] (-0.10,0.00) rectangle (0.10,0.55);
  }%
  \hspace{0.4em}%
}

\newcommand{\tabledownarrowred}{%
  \hspace{0.4em}%
  \tikz[baseline=0pt, x=1ex, y=1.7ex]{%
    \fill[red!70!black] (-0.42,0.45) -- (0,0.00) -- (0.42,0.45) -- cycle;
    \fill[red!70!black] (-0.10,0.45) rectangle (0.10,1.00);
  }%
  \hspace{0.4em}%
}

\usepackage[utf8]{inputenc}
\usepackage{booktabs}
\usepackage[hang,flushmargin]{footmisc}
\usepackage{threeparttable}
\usepackage{multirow}
\usepackage{rotating}
\usepackage[bibnewpage, nodoi]{apacite}
\usepackage[longnamesfirst, round]{natbib}
\usepackage{soul}
\usepackage{url}
\usepackage[colorlinks=false]{hyperref}
\usepackage{amsmath}
\usepackage{tikz}
\usepackage{pgfplots}
\pgfplotsset{width=10cm,compat=1.9}
\usepackage{setspace}
\usepackage{rotating}
\usepackage{color}
\usepackage{caption}

\usepackage[margin=1in]{geometry} 

\usepackage{ragged2e}

\usepackage{titlesec} 
\titleformat{\section}
  {\normalfont\Large\bfseries} 
  {\thesection} 
  {1em} 
  {}
\titleformat{\subsection}
  {\normalfont\large\bfseries} 
  {\thesubsection} 
  {1em} 
  {}

\usepackage{float}
\usepackage{caption}
\usepackage{makecell} 

\usepackage{graphicx} 
\usepackage{subcaption} 

\usepackage{amsmath, amssymb} 

\usepackage{enumitem}
\setlist[itemize]{noitemsep, topsep=0pt, left=1em}

\usepackage{hyperref}
\hypersetup{
    colorlinks=true,
    linkcolor=black,
    citecolor=black,
    urlcolor=black
}

\usepackage[american]{babel} 

\usepackage[toc,page]{appendix}
\usepackage{chngcntr} 
\usepackage[patch=none]{microtype}

\usepackage{titlesec}
\titlespacing*{\section}{0pt}{8pt}{4pt}
\titlespacing*{\subsection}{0pt}{6pt}{3pt}

\title{Suspense and Surprise in European Football\thanks{    We are grateful to Babatunde Buraimo, David Forrest, Ian McHale, and Juan de Dios Tena for sharing some of their data. An earlier version of this paper was presented at the annual conference of the European Sport Economics Association in Rotterdam, the Reading Online Sport Economics Seminar, and seminars hosted by the University of St. Gallen, the University of Liverpool, and Bielefeld University. We thank the participants at these events, as well as three anonymous referees and the editor, for their helpful comments and suggestions. The authors have no financial interest in the topic of this paper, and there are no conflicts of interest. Any remaining errors are our own.}}

\author{
\begin{tabular}{ccc}
\makecell{Raphael Flepp\footnote{Corresponding author. University of Zurich, Department of Business Administration. Email: raphael.flepp@business.uzh.ch} \\ \small \textit{University of Zurich}} & \makecell{Tim Pawlowski\footnote{University of Tübingen, Institute of Sports Science, LEAD Graduate School and Research Network, Interfaculty Research Institute for Sports and Physical Activity. Email: tim.pawlowski@uni-tuebingen.de} \\ \small \textit{University of Tübingen}} & \makecell{Travis Richardson\footnote{University of Tübingen, Institute of Sports Science. Email: travis-william.richardson@uni-tuebingen.de} \\ \small \textit{University of Tübingen}}  \\
\end{tabular}
}

\date{}

\date{July 2026}

\begin{document}
{\setstretch{1.0}\maketitle
\vspace{0.5mm}
  \begin{center}
    \small Author accepted manuscript
    \\ Published in the \textit{European Journal of Operational Research}
  \end{center}
\vspace{2em}}

\begin{abstract}
A key criterion for evaluating sports competitions is their attractiveness to consumers. In this paper, we propose using match-level suspense and surprise, which capture the entertainment utility generated by competitive balance and outcome uncertainty for sports spectators, to assess attractiveness. Through simulations, we derive a benchmark range for suspense and surprise based on an otherwise balanced match with home advantage, before analyzing more than 25,000 matches from Europe’s top five football leagues played between 2010/11 and 2023/24. Our findings reveal that mean match suspense is often lower than the benchmark range, particularly for top teams, while mean match surprise most often aligns with the benchmark. Moreover, we identify nuanced trends over time and highlight notable differences across leagues and clubs. These insights enhance our understanding of how the attractiveness of matches arises from competitive balance and carry important policy implications.

\noindent
\\
\vspace{2em}
\end{abstract}

\clearpage
\pagenumbering{arabic} 
\setcounter{page}{1}
\setstretch{1.5}
\section{Introduction}

Professional sports constitute an entertainment product, with many competitions and matches regularly attracting widespread attention. More generally, the attractiveness of a match in sports can be defined by its quality, its importance, or its competitive intensity \citep{Devriesere2025}. The latter is also referred to as outcome uncertainty, a concept rooted in the seminal works of \citet{Rottenberg1956} and \citet{Neale1964}, who argue that the unpredictability of competitive outcomes enhances fan welfare. Conceptually, competitive balance, which refers to the relative strengths of teams, translates into outcome uncertainty at various levels: a single match (short-term perspective), a sub-competition such as the championship race (mid-term perspective), or the dominance of a few teams over time (long-term perspective) \citep[e.g.,][]{Szymanski2003}. Subsequently, greater outcome uncertainty is assumed to lead to higher demand for sports. This relationship is known in the literature as the Uncertainty of Outcome Hypothesis (UOH), which primarily refers to the relation between outcome uncertainty and demand at the match level \citep[e.g.,][]{Fort2022, Pawlowski2018}.

Based on the premise that outcome uncertainty enhances entertainment value and is desirable for sports fans, league organizers frequently (re)design competitions, and policymakers justify market interventions such as salary caps, draft systems, or revenue redistribution schemes \citep[e.g.,][]{Fort1995}, which are prohibited or heavily restricted in most industries outside of professional sports \citep[e.g.,][]{Budzinski2012, Mehra2006}. Moreover, numerous studies investigate factors influencing outcome uncertainty, including competition design and tournament rules \citep[e.g.,][]{Scarf2009, Chater2021, Csato2022, Devriesere2025, Csato2026}, sporting rules \citep[e.g.,][]{Goller2020, Baker2022, Csato2023, DiMattia2023}, point-scoring systems \citep[e.g.,][]{Percy2009}, and scoring rates \citep[e.g.,][]{Scarf2019}. However, prior research has failed to provide consistent evidence that outcome uncertainty is a key driver of stadium attendance and TV viewership, particularly in European football \citep{Pawlowski2019, Collins2022, Johnson2022}. Accordingly, monitoring outcome uncertainty over time and treating it as a policy target appears misleading.

In this paper, we adopt a different perspective and propose suspense and surprise \citep{Ely2015} as measures for assessing match attractiveness and, when averaged across matches, the mean match attractiveness of certain teams and competitions. Both measures capture the entertainment utility that originates from the relative strengths of the teams in a match and unfolds through the resolution of information during the game. In this regard, suspense is a forward-looking concept that reflects the excitement or anxiety about future events and arises from the variance in beliefs about the outcome of the game in the next period. Surprise, in contrast, is a backward-looking concept that captures the emotional response to an unexpected event, resulting from the shift between the current and previous beliefs about the game's outcome \citep{Ely2015}.

A key reason for employing both measures is their clear link to sports demand. Building on the framework of \citet{Ely2015}, researchers have empirically investigated how viewers derive entertainment utility from suspense and surprise by approximating subjective beliefs with objective outcome probabilities and using within-game panel data, such as TV viewing figures \citep{Bizzozero2016, Buraimo2020, Kaplan2021, Richardson2023}, streaming data \citep{Simonov2023}, and social media activity \citep{Pawlowski2024}. A consistent finding across these studies is that within-game sports demand is driven by both suspense and surprise. 

Moreover, both measures are closely tied to outcome uncertainty, albeit in opposite ways: suspense tends to increase with higher uncertainty, whereas surprise is more likely to arise when an outcome departs from what was previously expected. Consequently, distinguishing between suspense and surprise is theoretically valuable, as it captures two distinct dimensions of entertainment utility within the broader concept of outcome uncertainty. This perspective may help explain the mixed empirical support for the UOH, as it suggests that the relationship between match outcome uncertainty and demand operates through at least two distinct mechanisms. Finally, the calculation of suspense and surprise follows a bottom-up approach to competitive balance and outcome uncertainty \citep{Basini2023}. This enables analyses at different levels of aggregation, such as the club or league level, thereby allowing for a more precise examination of patterns and trends.

We examine the status quo and the development of suspense and surprise in European football, the most extensively analyzed context for the UOH \citep{Collins2022}, and pose three questions: (i) How do match-level suspense and surprise depend on the (relative) strengths of the teams? (ii) How do suspense and surprise in European football matches compare to a theoretical benchmark based on the maximum attainable levels of suspense and surprise? (iii) How have suspense and surprise evolved over time? Addressing these questions enables league organizers and policymakers to make informed regulatory and league design decisions by considering the origins, levels, and trends of demand-relevant suspense and surprise.

We begin by analyzing how match-level suspense and surprise are influenced by the strengths of the competing teams. This is important because team-strength variation captures competitive balance \citep{Baker2022}. To investigate this, we simulate matches between teams with varying scoring rates, which reflect the underlying strengths of the competing teams \citep{Scarf2022}. We refer to a match as being ``perfectly balanced" when both teams have equal strength and their scoring rates just differ by home advantage. Our simulations reveal a key trade-off: suspense in a perfectly balanced match is higher when scoring rates are lower, whereas surprise increases with higher scoring rates. Based on this, we establish a benchmark range of suspense and surprise, derived from simulating perfectly balanced matches at varying scoring rates. In the second step, we derive empirical values of suspense and surprise from over 25,000 matches played in the top five European leagues: the English Premier League, the German Bundesliga, the Spanish La Liga, the Italian Serie A, and the French Ligue 1, covering the seasons from 2010/11 to 2023/24. 

Our findings suggest that mean suspense across all matches is lower than the benchmark range. This pattern is particularly evident in matches involving top teams, whereas matches involving non-top teams are relatively less likely to exhibit mean suspense levels significantly below the lower benchmark bound. In contrast, mean surprise values typically remain within the benchmark range. Regarding time trends, we find no evidence of a general decline in suspense and surprise across European football. Instead, the results point to heterogeneous patterns across leagues, including downward trends, upward trends, and cases with no clear trend. In addition, some league-level trends are driven primarily by matches involving one or two clubs.

With this study, we contribute to the literature on the attractiveness of sports competitions. In particular, our analysis is closely related to the recent work of \citet{Buraimo2022}, who demonstrate the relevance of match importance in terms of league outcomes for demand. In contrast, we focus on uncertainty surrounding the outcome of a specific match rather than on league outcomes and propose a measure of entertainment utility that is grounded in the relative strengths of the teams and evolves through the resolution of information during the game. As such, we argue that suspense and surprise provide complementary ways of assessing match attractiveness and, when averaged across matches, mean match attractiveness of certain teams and competitions.

The remainder of this paper is structured as follows: Section \ref{sec:generalsetup} conceptualizes suspense and surprise; Section \ref{sec:sim_matches} derives suspense and surprise from simulated matches involving teams of varying strength; Section \ref{sec:realm} analyzes suspense and surprise in real matches; and Section \ref{sec:dis_concl} presents the discussion and conclusion.

\section{General Set-up} \label{sec:generalsetup}

The procedures for determining suspense and surprise values in both simulated and real matches build on the conceptual framework proposed by \citet{Ely2015}. In this framework, suspense is induced by the variance in beliefs about the next period, whereas surprise arises from the change in beliefs between the previous and current periods. In the context of football, these beliefs pertain to the final outcome of a match and can be approximated with objective winning probabilities.

In their seminal work, \citet{Ely2015} develop the corresponding model in a highly general framework with finite state and signal spaces. Although their main contribution is theoretical and design-oriented, i.e., they study the optimal way in which information should be revealed over time, they also provide several empirical illustrations of information revelation and the resulting suspense and surprise processes in different settings, including tennis, European football, blackjack, and the Clinton--Obama Democratic Party presidential primary. In contrast to this general theoretical framework, however, all of these empirical illustrations are based on binary outcomes only, namely winning or losing. For European football, where draws constitute a common third outcome, they collapse the draw and away-win categories in their illustration.

\citet{Buraimo2020} were the first to operationalize the model while jointly accounting for all three possible outcomes, namely a home win, a draw, and an away win. Building on their approach, we define the backward-looking measure of surprise as follows:

\begin{equation}
\text{Surprise}_t = \sqrt{(p_t^H - p_{t-1}^H)^2 + (p_t^D - p_{t-1}^D)^2 + (p_t^A - p_{t-1}^A)^2}
\label{eq:surprise}
\end{equation}

\noindent where \( p_t \) refers to the probability of a home team win (\( p_{t}^{H} \)), a draw (\( p_{t}^{D} \)), or an away team win (\( p_{t}^{A} \)) at minute \( t \) in a given match.\footnote{While we use this specification of surprise in our analysis, it is worth mentioning that the correlation between this specification and an alternative specification that omits the probabilities of a draw is 99\%.} In contrast, suspense is a forward-looking measure that weights potential changes in outcome probabilities by the probabilities of either the home or away team scoring in the next minute, denoted by \( p_{t+1}^{HS} \) and \( p_{t+1}^{AS} \), respectively. Accordingly, we define suspense in a given minute \( t \) as follows: 

\begin{equation}
\text{Suspense}_{t} = \sqrt{\sum_{j \in \{H,D,A\}} \left( p_{t+1}^{HS} \left[ ( p_{t+1}^{j} | HS_{t+1}^{} ) - p_t^j \right]^2 + p_{t+1}^{AS} \left[ ( p_{t+1}^{j} | AS_{t+1}^{} ) - p_t^j \right]^2 \right)}
\label{eq:suspense}
\end{equation}

\noindent where \( \left( p_{t+1}^{j} \mid HS_{t+1}^{} \right) - p_t^j \) and \( \left( p_{t+1}^{j} \mid AS_{t+1}^{} \right) - p_t^j \) reflect the hypothetical change in outcome probabilities from minute \( t \) to \( t+1 \), given that either team scores in \( t+1 \).

To calculate the required (conditional) outcome probabilities on a minute-by-minute basis for each match, we follow the approach outlined by \citet{Buraimo2020} and implement an in-play model. For each game, we assume that a European football match resembles a ``Poisson match", where goals are scored independently by the home and away teams, following univariate Poisson distributions. The Poisson model has been suggested first by \citet{Maher1982} and is a classical choice for the distribution of the number of goals in football \citep[e.g.,][]{Csato2023, Scarf2019, Scarf2022, vanEetvelde2019}. Specifically, the number of goals scored by the home team, \( X \sim \text{Poisson}(\lambda_H) \), and the number of goals scored by the away team, \( Y \sim \text{Poisson}(\lambda_A) \). We distribute the team-specific scoring rates, \( \lambda_H \) and \( \lambda_A \), across the match using historical, league-specific minute-by-minute empirical goal distributions to derive time-specific scoring rates. These minute-by-minute goal-scoring rates are combined with the timing of major in-game events, namely goals scored and red cards received, to estimate the outcome probabilities for each match at any given minute. Specifically, scoring in a given minute is simulated by drawing from a Bernoulli distribution, where the success probability is set to the estimated minute-by-minute scoring rate. That is, \( G_t \sim \text{Bernoulli}(p_t^{S}) \), where \( G_t \) represents whether a goal is scored in minute \( t \), and \( p_t^{S} \) denotes the estimated team-specific goal-scoring probability in that minute \citep{Chater2021}. The simulation exercise is repeated 100,000 times for each minute, resulting in 9 million simulated match-minutes. 

One important consideration in this modeling process is the adjustment of scoring rates following red cards. \citet{Vecer2009} estimate the effect of a red card on scoring rates using betting data from matches played at the 2006 FIFA World Cup and UEFA EURO 2008. They find that, for the remainder of the match, the scoring rate of the team receiving a red card is reduced, on average, to $\frac{2}{3}$ of its original value, while that of the opposing team increases to 1.2 times its original value. We adopt these parameters in our modeling process. 

However, because \citet{Vecer2009} use relatively old data from a small sample of national-team matches, we conducted robustness checks to assess whether our results are sensitive to alternative parameter values. To do so, we first selected, from the real-match data described in Section \ref{sec:realm}, the subsample of matches in which at least one red card was awarded. Using this subsample, we conducted sensitivity analyses with scoring-rate multipliers of 0.5, 0.66, and 0.8 for the team receiving the red card, and 1.1, 1.2, and 1.3 for the opposing team. Overall, our findings did not change materially. We therefore decided to retain the ratios originally proposed by \citet{Vecer2009} in our analysis.

A second important consideration in the modeling process is the treatment of injury time. Injury time is regularly added in both the first and second half, resulting in substantial variation in effective playing time across matches. In our modeling approach, we therefore introduce one additional injury-time minute in each half, namely 45+1 and 90+1, into which all events occurring during injury time are collapsed. While this remains an approximation, it allows us to distinguish more clearly between regular playing time and injury time in the analysis. Adding further injury-time minutes in the first half is problematic because injury time there is usually short. Likewise, introducing multiple additional injury-time minutes in the second half would be somewhat arbitrary and would not materially improve our approach, as we cannot account for heterogeneity in injury-time length across matches in the simulation.
 
The resulting outcome probabilities from this modeling process are used to calculate suspense and surprise on a minute-by-minute basis, which are then aggregated over all minutes at the match level. Note, since we are missing a value for suspense in minute 90+1 (as suspense is forward-looking), we effectively use 91 minute-suspense values for calculating aggregated suspense. Moreover, surprise in minute 1 is always zero unless a goal or red card happened in minute 1, in which case we calculated surprise using the closing odds as reference category. As such, we effectively use 92 minute-surprise values for calculating aggregated surprise. Overall, this procedure yields one suspense and one surprise value per match.

\section{Simulated Matches}\label{sec:sim_matches}

To investigate how suspense and surprise are influenced by team strengths, we simulate matches between teams with varying scoring rates. We consider 1,326 unique scoring rate combinations, covering all pairings of \( \lambda_H \) and \( \lambda_A \) ranging from 0 and 5 in increments of 0.1. Since football is a low-scoring sport, with a mean scoring rate of approximately 1.5 goals per team per match \citep[e.g.,][]{Baker2018, Scarf2022}, a scoring rate of 5 represents a really high value. 

For each pairing, we simulate 100,000 matches, resulting in a total of 132,600,000 hypothetical matches. The simulations use historical minute-by-minute empirical goal distributions from the English Premier League to determine the minute-by-minute scoring rates.\footnote{The empirical goal distributions across the top five European leagues are similar, and we do not expect that selecting the empirical goal distribution of another top five league would significantly affect our results.} Red cards are incorporated into the simulation framework in a manner identical to how goals are simulated. Using the historical data, we calculate the empirical minute-by-minute probability of a red card occurring separately for home and away teams across all 92 minutes (i.e., including minutes 45+1 and 90+1 as described before). In each simulated game, for every minute we draw from a Bernoulli distribution using these empirical probabilities to determine whether a red card occurs at that minute for each team. Once a red card is assigned to a team, its scoring rate is reduced by a factor of $\frac{2}{3}$, while the opposing team's scoring rate is increased by a factor of 1.2 for all remaining minutes of the match. This red card state is tracked cumulatively so that if a team accumulates more red cards than the opposing team at any point, the scoring rate adjustments are applied from that minute onward for the remainder of the game.

\begin{figure}[ht!]
\vspace{5mm}
\centering
\caption{\centering Suspense and Surprise as a Function of Teams’ Scoring Rates}
    \label{fig:simulation_graph}
    \begin{subfigure}{0.49\textwidth}
        \centering
        \includegraphics[width=\linewidth,trim = 20mm 20mm 10mm 20mm, clip]{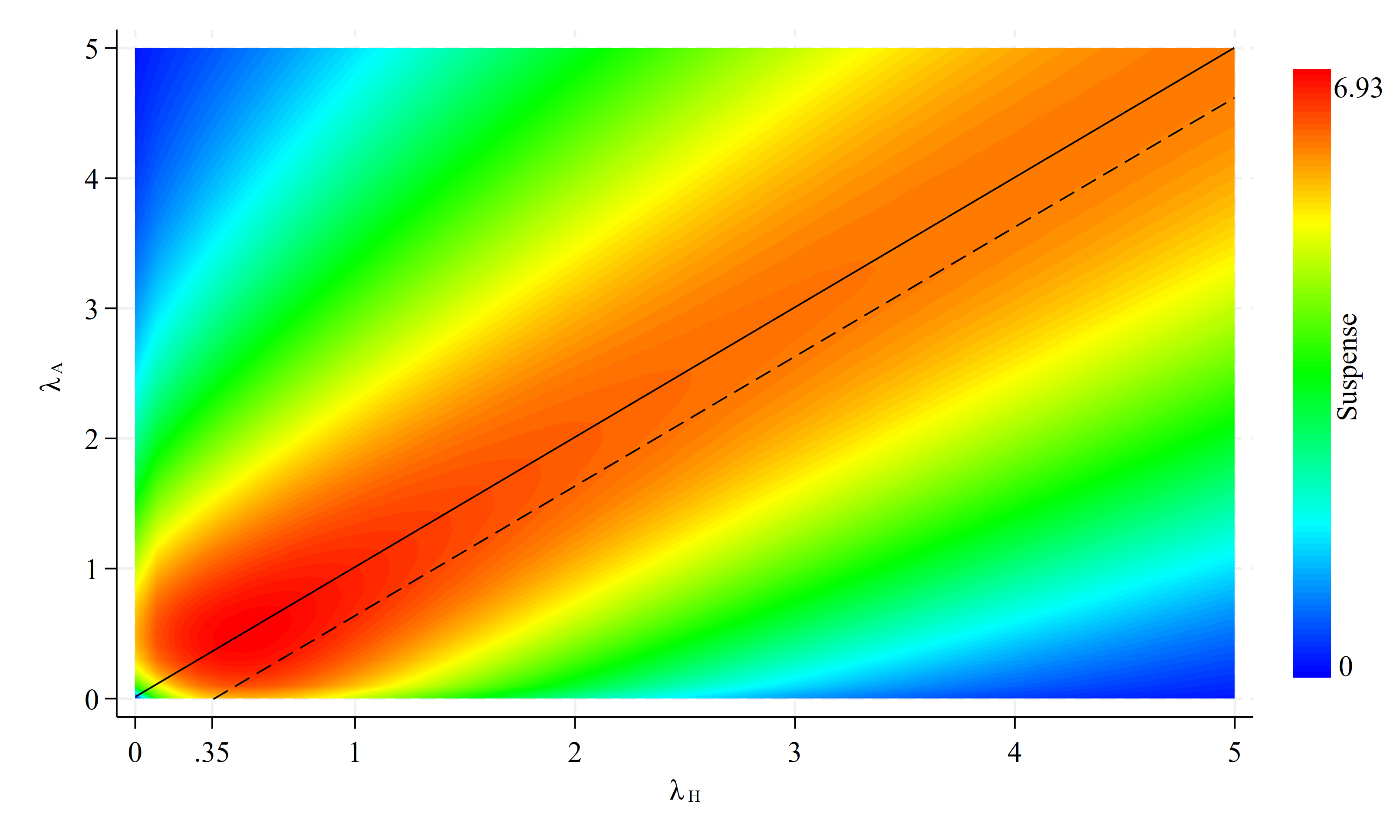}
        \caption{Suspense}
        \label{fig:sus}
    \end{subfigure}
    \hfill
    \begin{subfigure}{0.49\textwidth}
        \centering
        \includegraphics[width=\linewidth,trim=20mm 20mm 10mm 20mm, clip]{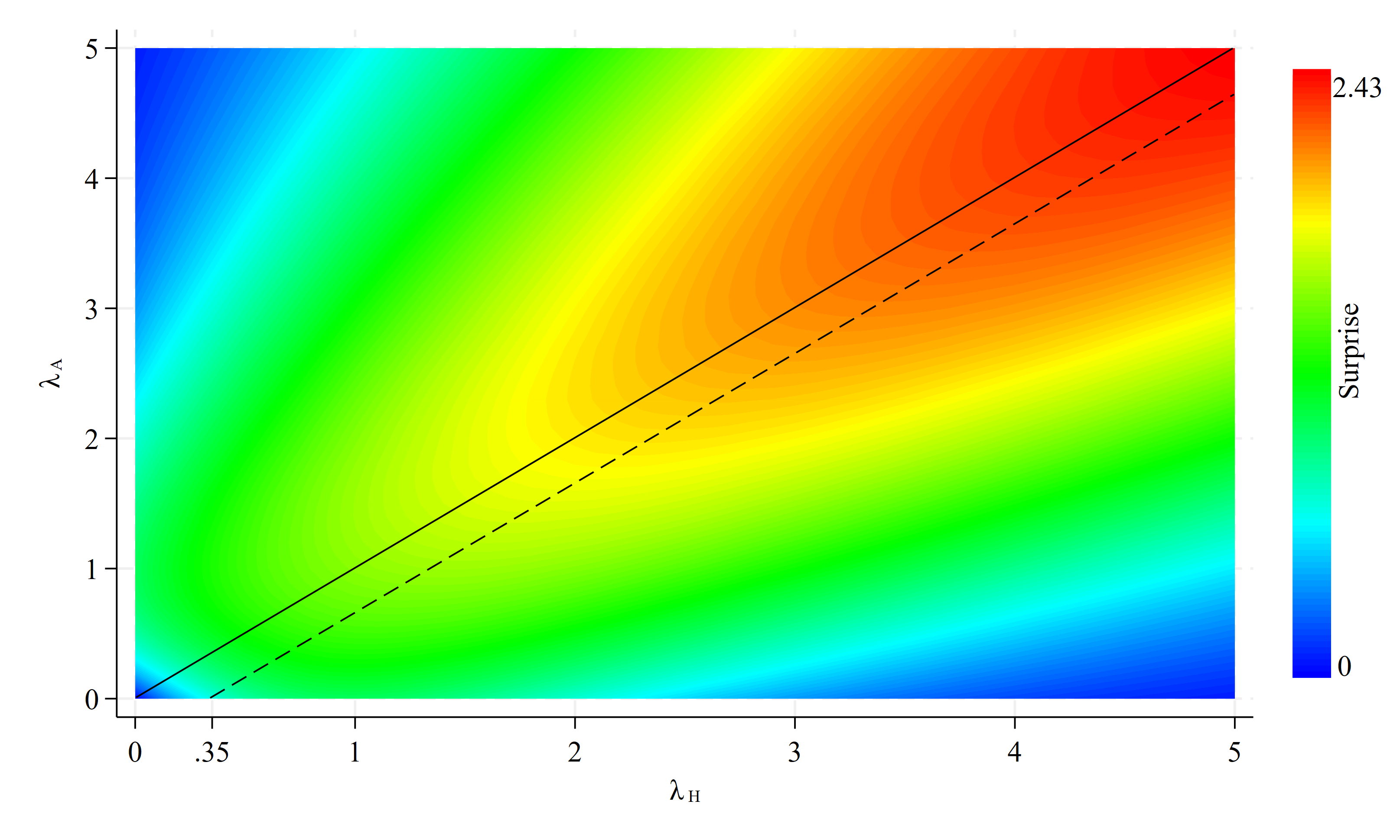}
        \caption{Surprise}
        \label{fig:sur}
    \end{subfigure}
    \caption*{\footnotesize\textit{Note:} This figure plots match-level suspense (Figure \ref{fig:sus}) and surprise (Figure \ref{fig:sur}) values as a function of the scoring rates $\lambda_H$ for the home team and $\lambda_A$ for the away team, assuming a Poisson match where goals are scored independently following univariate Poisson distributions. The 45-degree line reflects match-level suspense and surprise in matches between teams of equal strength and with equal scoring rates. The parallel shifted dashed line reflects the corresponding values for otherwise balanced matches with a home advantage of 0.35 goals.}
\end{figure}

Subsequently, we calculate the suspense and surprise values for each simulated match, following the methodology outlined in Section \ref{sec:generalsetup}. Figure \ref{fig:simulation_graph} presents the simulation results for the mean level of suspense (Figure \ref{fig:sus}) and surprise (Figure \ref{fig:sur}) across different combinations of scoring rates, \( \lambda_H \) and \( \lambda_A \). Suspense is low (indicated in blue) when scoring rates are highly unequal, as one team is overwhelmingly dominant and almost certain to win the match. Conversely, suspense tends to be high (indicated in red) when scoring rates are equal. However, suspense is zero when \( \lambda_H = \lambda_A = 0 \), since the outcome of a draw is certain, even though the match is perfectly competitively balanced. This illustrates the fundamental distinction between competitive balance and outcome uncertainty, as also noted by \citet{Scarf2019}.

Importantly, even in a perfectly balanced match, the level of suspense depends on the absolute values of the scoring rates. When both rates are close to zero, the probability of a draw is high, as neither team is likely to score. As the scoring rates increase together while remaining equal, the likelihood of a draw decreases, and the probability of either a home or away win increases \citep{Scarf2022}. Therefore, different levels of equal scoring rates translate into different levels of suspense. Suspense increases for \( 0 < \lambda_H = \lambda_A < 0.5 \), reaches a maximum at \( \lambda_H = \lambda_A = 0.5 \), and decreases again for \( \lambda_H = \lambda_A > 0.5 \), as the two rates continue to rise together. This finding adds nuance to the conjectures of \citet{Scarf2019}, who suggest that suspense strictly decreases as scoring rates increase.

Interestingly, surprise follows the opposite pattern, increasing strictly with higher and more equal scoring rates. It is zero when \( \lambda_H = \lambda_A = 0 \), as no goals are expected and a draw is certain. In our simulations, the maximum level of surprise occurs at \( \lambda_H = \lambda_A = 5 \). Because these are the highest scoring rates considered in our simulations, this should be interpreted as a local maximum within the simulated range. Remarkably, higher but unequal scoring rates can produce levels of surprise similar to those observed under lower but equal scoring rates. For example, the surprise generated in a match with \( \lambda_H = \lambda_A = 1 \) is similar to that in a match with \( \lambda_H = 3 \) and \( \lambda_A = 1.5 \). 

These findings contradict the conjecture by \citet{Scarf2019}, who suggest that higher scoring rates always reduce surprise. One possible explanation for this discrepancy is the simpler definition and operationalization of surprise used by \citet{Scarf2019}, who classify a match as surprising if a team defeats another team whose win percentage is at least 20 percentage points higher, and as non-surprising otherwise. Their measure therefore captures only whether the final outcome was surprising, but not the within-match variations in outcome probability that contribute to surprise as originally defined by \citet{Ely2015}.

More broadly, the simulation results in Figure \ref{fig:simulation_graph} highlight a trade-off between suspense and surprise when relative team strength is held constant. This pattern is clearest in perfectly balanced matches: suspense is higher at lower scoring rates, except when rates are close to zero, whereas surprise is higher at higher scoring rates. Intuitively, lower scoring rates tend to keep matches closer for longer, which increases suspense, while higher scoring rates generate more scoring events, creating more opportunities for surprising developments. Thus, for a given degree of team-strength imbalance, changes in the overall scoring environment can move suspense and surprise in opposite directions.

We further use the simulated matches to examine how suspense and surprise respond to changes in team-strength imbalance. We define team-strength imbalance as the absolute log difference in the two teams’ scoring rates, $|\log(\lambda_H)-\log(\lambda_A)|$, and control for the average log scoring-rate level, $[\log(\lambda_H) +\log(\lambda_A)]/2$. We then separately regress mean suspense and mean surprise on team-strength imbalance and the average log scoring-rate level. The standardized coefficients on team-strength imbalance show that, conditional on the average scoring-rate level, suspense responds more strongly to imbalance than surprise. To illustrate, a one-standard-deviation increase in team-strength imbalance is associated with a 1.12-standard-deviation decrease in suspense, compared with a 0.78-standard-deviation decrease in surprise. Thus, for the same increase in imbalance, the decline in suspense is about 44\% larger than the decline in surprise.

To contextualize the current empirical values of suspense and surprise in European football (see Section \ref{sec:realm}), we use the simulation results to derive a benchmark range for both metrics for otherwise balanced matches with home advantage and varying absolute scoring rates. Under equal playing strength, the expected scoring differential between the home and away team should reflect home advantage in the standard double round-robin setting of the leagues under consideration. Using the empirical distribution of the mean number of goals scored per match by home and away teams, we estimate that home advantage corresponds, on average, to an increase of 0.35 in the home team’s scoring rate. This home advantage is illustrated by a parallel shift of the 45-degree line in Figure \ref{fig:simulation_graph}.

Since the scoring-rate levels that would prevail in a perfectly balanced match are not known ex ante, we choose a deliberately broad range of low and high scoring rates. For the low-scoring case, we choose a scoring rate of 0.5, which corresponds to the 10th percentile of the empirical distribution of the mean number of goals scored per team per match in our sample. Accordingly, when incorporating home advantage, we set \( \lambda_H = 0.85 \) and \( \lambda_A = 0.5 \). Likewise, for the high-scoring case, we choose a scoring rate of 2.5, which corresponds to the 90th percentile of the empirical distribution in our sample.\footnote{As a sensitivity check, we also use the 15th and 85th percentiles to define the low- and high-scoring-rate scenarios. The results remain qualitatively unchanged. We interpret this as a meaningful sensitivity check, while acknowledging that whether empirical suspense and surprise values fall below, within, or above the benchmark range necessarily depends on the width of that range. Much narrower percentile intervals would mechanically tighten the benchmark bands and could lead to more empirical values to fall outside them.} Taking home advantage into account, we set \( \lambda_H = 2.85 \) and \( \lambda_A = 2.5 \).\footnote{We additionally test whether home advantage varies with scoring rates using league-season observations. Specifically, we regress average home-away goal difference on mean-centered average total goals per match, weighted by the number of matches. The estimated relationship is small and statistically insignificant ($\beta=-0.036$, $p=0.510$), indicating no systematic variation in home advantage across scoring rates. We also test whether home advantage differs across leagues. Pairwise comparisons show that La Liga has a somewhat higher average home-away goal difference (0.42) than several other leagues, significantly exceeding Ligue 1, the Premier League, and Serie A, but not the Bundesliga. Recalculating the La Liga benchmark bands using a home-advantage adjustment of 0.42 instead of 0.35 leaves the results virtually unchanged.}

Note that, in contrast to our approach, the literature often identifies unique upper and lower bounds of competitive balance indices based on the theoretically most equal or most unequal distribution of wins or points \citep[e.g.,][]{AvilaCano2021, AvilaCano2023, Depken1999, Owen2007}. Alternatively, realized concentrations are compared with established thresholds for market concentration set by authorities such as the European Commission or the U.S. Federal Trade Commission \citep[e.g.,][]{AvilaCanoTrigueroRuiz2023}. In addition, we derive the benchmark values endogenously from the premise that viewers value suspense and surprise. Under this premise, suspense is highest when scoring rates are low and equal (i.e., both equal to 0.5), whereas surprise is highest when scoring rates are high but still equal.

Table \ref{tab:descriptive_statistics_sim} presents the descriptive statistics for suspense and surprise in otherwise perfectly balanced matches with home advantage and low (\( \lambda_H = 0.85, \ \lambda_A = 0.5 \)) or high scoring rates (\( \lambda_H = 2.85, \ \lambda_A = 2.5 \)). Within this benchmark range, suspense has a lower bound of 6.08 and an upper bound of 6.70, while surprise ranges from 1.34 to 1.99. These values define the benchmark range boundaries against which we compare the empirical values of suspense and surprise.

\begin{table}[ht]
\vspace*{2mm}
    \centering
    \caption{Descriptive Statistics for Otherwise Balanced Matches with Home Advantage}
    \label{tab:descriptive_statistics_sim}
    \small
    \begin{tabular}{l l c c c c c c}
        \toprule
         &  & N & Mean & Median & SD & Min & Max \\
        \midrule
        \multirow{2}{*}{\makecell[tl]{Low Scoring Rates \\(\( \lambda_H = 0.85; \lambda_A = 0.5 \))}  }
        & Suspense & 100,000 & 6.70 & 7.15 & 1.68 & 0.24 & 9.31 \\
        & Surprise & 100,000 & 1.34 & 1.11 & 0.68 & 0.62 & 6.68 \\
        \midrule
        \multirow{2}{*}{\makecell[tl]{High Scoring Rates\\ (\( \lambda_H = 2.85; \lambda_A = 2.5 \))}  }
        & Suspense & 100,000 & 6.08 & 6.49 & 1.87 & 0.46 & 9.39 \\
        & Surprise & 100,000 & 1.99 & 1.84 & 0.85 & 0.41 &  6.70 \\
        \bottomrule
    \end{tabular}
    \caption*{\footnotesize\textit{Notes:} Suspense and surprise in simulated, otherwise balanced matches with a home advantage of 0.35 in $\lambda_H$, under low-scoring conditions ($\lambda_H = 0.85$, $\lambda_A = 0.5$) and high-scoring conditions ($\lambda_H = 2.85$, $\lambda_A = 2.5$).}
\end{table}

\section{Real Matches}\label{sec:realm}
\subsection{Data and Sample}

Our empirical dataset comprises over 25,000 men's football matches played in the top five European leagues, i.e., the English Premier League, the German Bundesliga, the Spanish La Liga, the Italian Serie A, and the French Ligue 1, between the 2010/11 and 2023/24 seasons. Depending on the league, the dataset includes between 4,284 and 5,320 matches per league. Match information, including the timing of goals and red cards, was collected from fbref.com, while pre-match closing odds and over-under totals were obtained from oddsportal.com.

To estimate the scoring rates \( \lambda_H \) and \( \lambda_A \) for each game, we follow the methodology of \citet{Buraimo2020}, which relies on betting odds and uses an optimization procedure based on the assumption that goals scored follow Poisson distributions. Betting odds have been shown to contain highly accurate information on football matches \citep{Wunderlich2025} and to have higher predictive quality than alternative approaches to measuring team strength, such as Elo ratings \citep{Hvattum2010, Wunderlich2018}.

We adjust the pre-match betting odds for the overround to derive implied probabilities for a home win, draw, and away win. While these closing odds already capture some information about the expected number of goals, since a team more likely to win is also more likely to score more, over/under (O/U) odds provide additional, independent constraints on the distribution of total goals scored. Specifically, O/U odds for thresholds ranging from 0.5 to 5.5 goals are converted into implied probabilities and incorporated into the same optimization procedure, ensuring that the selected scoring rates not only match expected match outcomes but also align with market expectations about total goals. Using the procedure outlined in Section \ref{sec:generalsetup}, we calculate match-level suspense and surprise values for all 25,389 matches played.

To validate our assumption that match-level suspense and surprise are positively related to sports demand, we regress both measures on TV audience using the original dataset from \citet{Buraimo2022}. This dataset comprises 790 English Premier League matches televised between the second half of the 2013/14 season and the end of the 2018/19 season. In Columns 2 and 3 of Table \ref{tab:tv_demand} in Appendix A, we show that suspense and surprise are positively associated with TV demand and help explain demand over and above league-outcome-specific measures proposed by \citet{Buraimo2022}, namely match significance with respect to the championship, European qualification, and relegation. A detailed description of the analysis and results is provided in Appendix \ref{sec:tvdemand}.

\begin{table}[ht!]
    \singlespacing
    \centering
    \caption{Descriptive Statistics of Suspense and Surprise in Men's European Football Matches}
    \label{tab:descriptive_statistics}
    \small
    \begin{tabular}{l l c c c c c c c c}
        \toprule
        League & Metric & N & Mean & Median & SD & Min & Max & BM & $<$ BM \\
        \midrule
        \multirow{2}{*}{Top 5 Leagues} 
        & Suspense & 25,389 & 5.90 & 6.59 & 2.15 & 0.08 & 9.31 & 6.08 & *** \\
        & Surprise & 25,389 & 1.42 & 1.80 & 0.81 & 0.12 & 5.91 & 1.34 &  \\
        \midrule
        \multirow{2}{*}{Premier League} 
        & Suspense & 5,320 & 5.78 & 6.43 & 2.19 & 0.08 & 9.17 & 6.08 &  ***  \\
        & Surprise & 5,320 & 1.41 & 1.17 & 0.82 & 0.14 & 5.36 & 1.34 &  \\
        \midrule
        \multirow{2}{*}{Bundesliga}  
        & Suspense & 4,284 & 5.85 & 6.47 & 2.19 & 0.26 & 9.11 & 6.08 &  ***  \\
        & Surprise & 4,284 & 1.44 & 1.27 & 0.80 & 0.14 & 5.91 & 1.34 &  \\
        \midrule
        \multirow{2}{*}{La Liga}  
        & Suspense & 5,320 & 5.89 & 6.64 & 2.22 & 0.16 & 9.26 & 6.08 & *** \\
        & Surprise & 5,320 & 1.41 & 1.16 & 0.82 & 0.12 & 5.69 & 1.34 &  \\
        \midrule
        \multirow{2}{*}{Serie A}  
        & Suspense & 5,320 & 5.92 & 6.60 & 2.13 & 0.23 & 9.15 & 6.08 &  *** \\
        & Surprise & 5,320 & 1.43 & 1.22 & 0.82 & 0.14 & 5.44 & 1.34 &  \\
        \midrule
        \multirow{2}{*}{Ligue 1}  
        & Suspense & 5,145 & 6.05 & 6.70 & 2.04 & 0.17 & 9.31 & 6.08 &  \\
        & Surprise & 5,145 & 1.42 & 1.20 & 0.79 & 0.12 & 5.17 & 1.34 &  \\
        \bottomrule
    \end{tabular}
    \caption*{\footnotesize\textit{Notes}: Suspense and surprise values are calculated as described in Section \ref{sec:generalsetup}. Fewer games are observed for Ligue 1 due to the unbalanced schedule during the COVID-19 pandemic and the league's reduction from 20 to 18 teams for the 2023/24 season. BM refers to the lower bound of the benchmark range. $<$ BM indicates a one-sided t-test assessing whether the mean is below the benchmark. *** $p < 0.01$, ** $p < 0.05$, * $p < 0.1$.}
\end{table}

Table \ref{tab:descriptive_statistics} presents descriptive statistics for suspense and surprise across leagues. League comparisons reveal several differences, with mean suspense ranging from 5.78 in the Premier League to 6.05 in Ligue 1 over the observation period. Similarly, mean surprise varies from 1.41 in the Premier League and La Liga to 1.44 in the Bundesliga. Notably, the relatively large standard deviations indicate substantial variation over time and across matches. The last column of Table \ref{tab:descriptive_statistics} indicates whether mean suspense and surprise values fall significantly below the lower bound of the benchmark range for perfectly balanced matches as defined in Section \ref{sec:sim_matches}, based on a one-sided t-test. The results show that mean suspense values are significantly lower than the benchmark’s lower bound for all leagues but Ligue 1. In contrast, mean surprise values across all five leagues remain above the lower bound of the benchmark range. This suggests that, while mean suspense in matches from the top five European leagues is comparatively low, mean surprise remains within the benchmark range. As a robustness check, we excluded the COVID-19-affected seasons 2019/20, 2020/21, and 2021/22 from our sample, and the results remained similar.

In the following analysis, we exploit the granularity of match-level suspense and surprise to examine trends within each league individually. Importantly, however, while we report observed trends, we cannot draw conclusions about the underlying mechanisms, whether COVID-19-related or otherwise.

\subsection{Suspense and Surprise Trends in the English Premier League}

\begin{figure}[ht!]
    \centering
    \caption{Suspense and Surprise in the English Premier League}
    \label{fig:suspense_surprise_epl}
    \begin{subfigure}{\textwidth}
        \centering
        \includegraphics[width=\linewidth,trim = 0mm 0mm 0mm 0mm, clip]{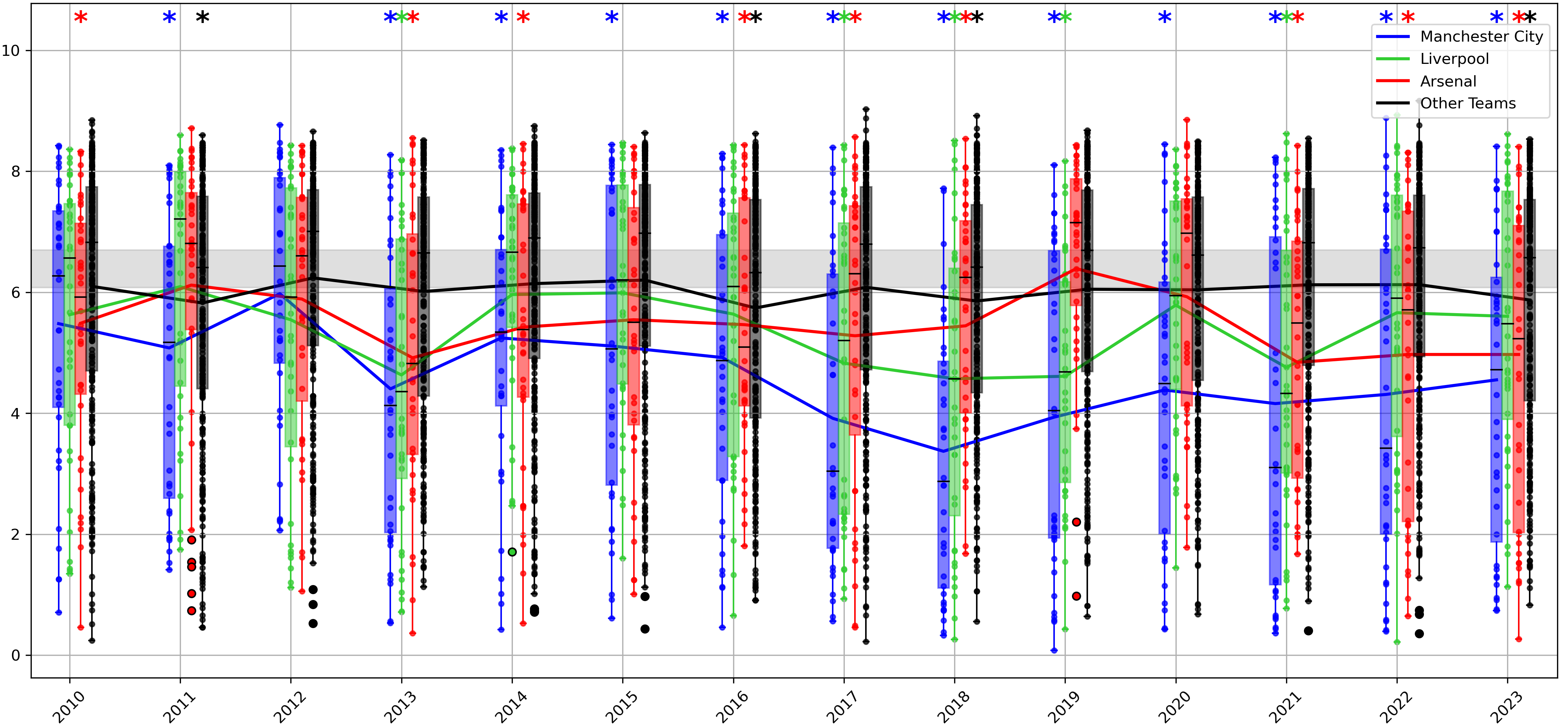}
        \caption{Suspense}
        \label{fig:suspense_epl}
    \end{subfigure}
    
    \begin{subfigure}{\textwidth}
        \centering
        \includegraphics[width=\textwidth,trim = 0mm 0mm 0mm 0mm, clip]{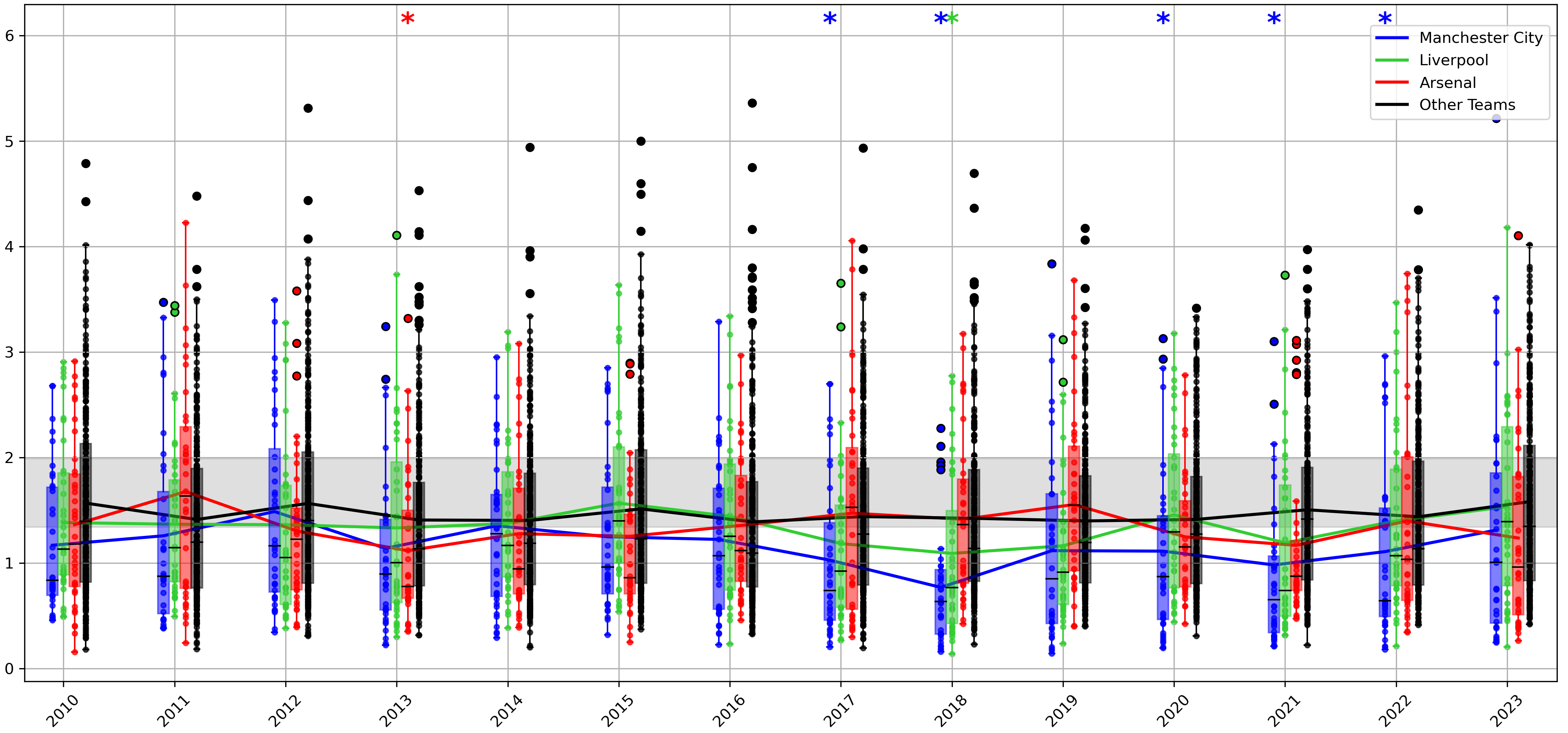}
        \caption{Surprise}
        \label{fig:surprise_epl}
    \end{subfigure}
    \caption*{\footnotesize\textit{Notes}: This figure displays box plots for mean suspense (Figure \ref{fig:suspense_epl}) and mean surprise (Figure \ref{fig:surprise_epl}) per game across teams and seasons in the English Premier League. The grey horizontal band represents the benchmark range as defined in Section \ref{sec:sim_matches}. For suspense, the upper (lower) bound of the benchmark range corresponds to  mean suspense of an otherwise balanced match with home advantage and low (high) scoring rates of \( \lambda_H = 0.85 \) and \( \lambda_A = 0.5 \) (\( \lambda_H = 2.85, \ \lambda_A = 2.5 \)). For surprise, the upper (lower) bound of the benchmark range corresponds to mean surprise of an otherwise balanced match with home advantage and high (low) scoring rates of \( \lambda_H = 2.85 \) and \( \lambda_A = 2.5 \) (\( \lambda_H = 0.85, \ \lambda_A = 0.5 \)). * indicates a significant difference (\( p \) $< 0.05$) between the mean value at the team-season level and the lower bound of the benchmark range, based on a one-sided t-test.}
\end{figure}

We begin by examining the development of suspense and surprise in the English Premier League. Figure \ref{fig:suspense_surprise_epl} presents box plots showing mean suspense and surprise per game across teams and seasons. We visualize the three top teams separately, defined as those that accumulated the most points during our observation period from 2010/11 to 2023/24, while grouping all remaining teams together. According to this definition, the top three teams in the Premier League were Manchester City, Liverpool, and Arsenal. The grey horizontal band represents the benchmark range for a perfectly balanced match as defined in Section \ref{sec:sim_matches}. An asterisk (*) indicates whether mean suspense or surprise values for a team in a given season are significantly below the lower bound of the benchmark range (\( p \) $< 0.05$), based on a one-sided t-test. 

Figure \ref{fig:suspense_epl} shows that mean suspense values for Manchester City matches tend to be the lowest and are almost always significantly below the lower benchmark bound. For Liverpool and Arsenal, a similar pattern emerges, though it is less pronounced. While visual inspection of Figure \ref{fig:suspense_epl} suggests that mean suspense for all other teams only rarely falls below the benchmark range, Panel A of Table \ref{tab:epl_suspense_cond} in Appendix \ref{sec:team_seasons} shows that matches involving Chelsea, Manchester United, and Tottenham Hotspur also recorded multiple seasons with mean suspense values below the lower benchmark band. In addition, the 2018/19 season stands out, as 12 of the 20 teams exhibited low suspense values. With regard to surprise, Figure \ref{fig:surprise_epl} and Panel B of Table \ref{tab:epl_suspense_cond} in Appendix \ref{sec:team_seasons} show that mean values less frequently fall below the benchmark range. Manchester City again stands out as an exception, with values falling significantly below the benchmark range in five seasons during our observation window.

Remarkably, Panels A and B of Table \ref{tab:epl_suspense_cond} show that comparatively low levels of suspense and/or surprise are generated not only by particularly strong teams, but also by particularly weak teams in a given season. Examples include Huddersfield in 2018/19 (finishing the season with 16 points), Norwich in 2019/20 and 2021/22 (21 and 22 points, respectively), and Sheffield United in 2023/24 (16 points). Since such teams are typically relegated on a regular basis, however, we do not present them in Figure \ref{fig:suspense_epl}. Note also that Leicester City’s Premier League title in the 2015/16 season, which has been characterized as an exceptionally unlikely sporting success \citep{BBC2016}, is likewise reflected in our data. In that season, the mean values of suspense and surprise for Leicester City matches (suspense: 6.23; surprise: 1.51) were higher than those for all other Premier League teams excluding Leicester City (suspense: 5.99; surprise: 1.46).

To formally investigate the magnitude of time trends in suspense and surprise, we estimate linear regressions using the natural log of suspense and surprise as the dependent variable and season year as a continuous independent variable.\footnote{Since our trend analysis is conducted at the match-level across multiple seasons, implementing a time series approach \citep[e.g.,][]{Lee2012} is not feasible. To account for dependence within match-pair observations, we cluster standard errors at the home team-away team level.} Table \ref{tab:time_trends_epl} summarizes the results. Column (1) indicates a slight decline in suspense over time, while Column (2) reveals that this negative trend is driven entirely by matches involving Manchester City, with suspense in those matches decreasing by approximately 3.8\% per season. Columns (3) and (4) show a similar pattern for surprise: matches involving Manchester City exhibit an additional decline of approximately 2.7\% per season relative to the baseline trend. By contrast, matches involving Arsenal, Liverpool, or the remaining teams do not show a comparable decline.

\begin{table}[ht]
    \centering
    \caption{English Premier League Time Trends}
    \label{tab:time_trends_epl}
    \small
    \begin{tabular}{l c c c c}
        \toprule
        & \multicolumn{2}{c}{Ln(suspense)} & \multicolumn{2}{c}{Ln(surprise)} \\
        \cmidrule(lr){2-3} \cmidrule(lr){4-5}
        & (1) & (2) & (3) & (4) \\
        \midrule
        Season & $-$0.005** & $-$0.000 & $-$0.003 & 0.001 \\
               & (0.002)  & (0.002)  & (0.002)  & (0.002)  \\
        Manchester City × season &  & $-$0.038*** &  & $-$0.027*** \\
                                 &  & (0.009)   &  & (0.009)   \\
        Liverpool × season &  & $-$0.005 &  & $-$0.006 \\
                                   &  & (0.006)  &  & (0.007)  \\
        Arsenal × season &  & $-$0.008 &  & $-$0.006 \\
                         &  & (0.008) &  & (0.007) \\
        \midrule
        Main effects &  &Yes  &  &Yes  \\
        Observations & 5,320 & 5,320 & 5,320 & 5,320 \\
        Adj-R\textsuperscript{2}  & 0.001 & 0.055 & 0.000 & 0.030 \\
        \bottomrule
    \end{tabular}
    \caption*{\footnotesize\textit{Notes}: This table reports the results from OLS regressions with robust standard errors, clustered at the home team–away team level, shown in parentheses. Season is defined as a continuous variable. **** $p < 0.01$, ** $p < 0.05$.}
\end{table}

Overall, these patterns align with Manchester City's dominance, having won seven league titles during our sample period. At the same time, however, suspense and surprise values in matches involving other Premier League teams remained relatively unchanged over time.

\subsection{Suspense and Surprise Trends in the German Bundesliga}

\begin{figure}[ht!]
    \centering
    \caption{Suspense and Surprise in the German Bundesliga}
    \label{fig:suspense_surprise_gbl}
    \begin{subfigure}{\textwidth}
        \centering
        \includegraphics[width=\linewidth,trim = 0mm 0mm 0mm 0mm, clip]{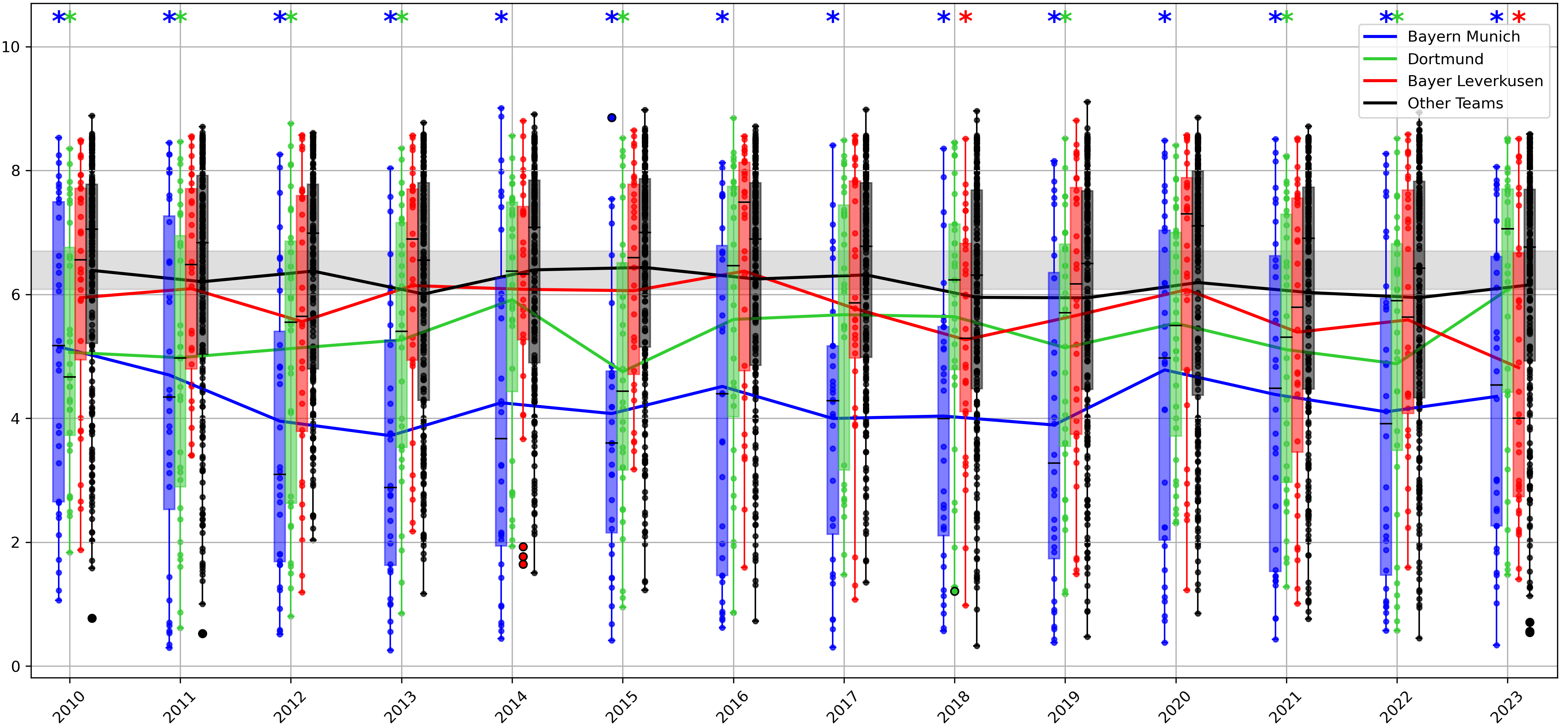}
        \caption{Suspense}
        \label{fig:suspense_gbl}
    \end{subfigure}
    
    \begin{subfigure}{\textwidth}
        \centering
        \includegraphics[width=\textwidth,trim = 0mm 0mm 0mm 0mm, clip]{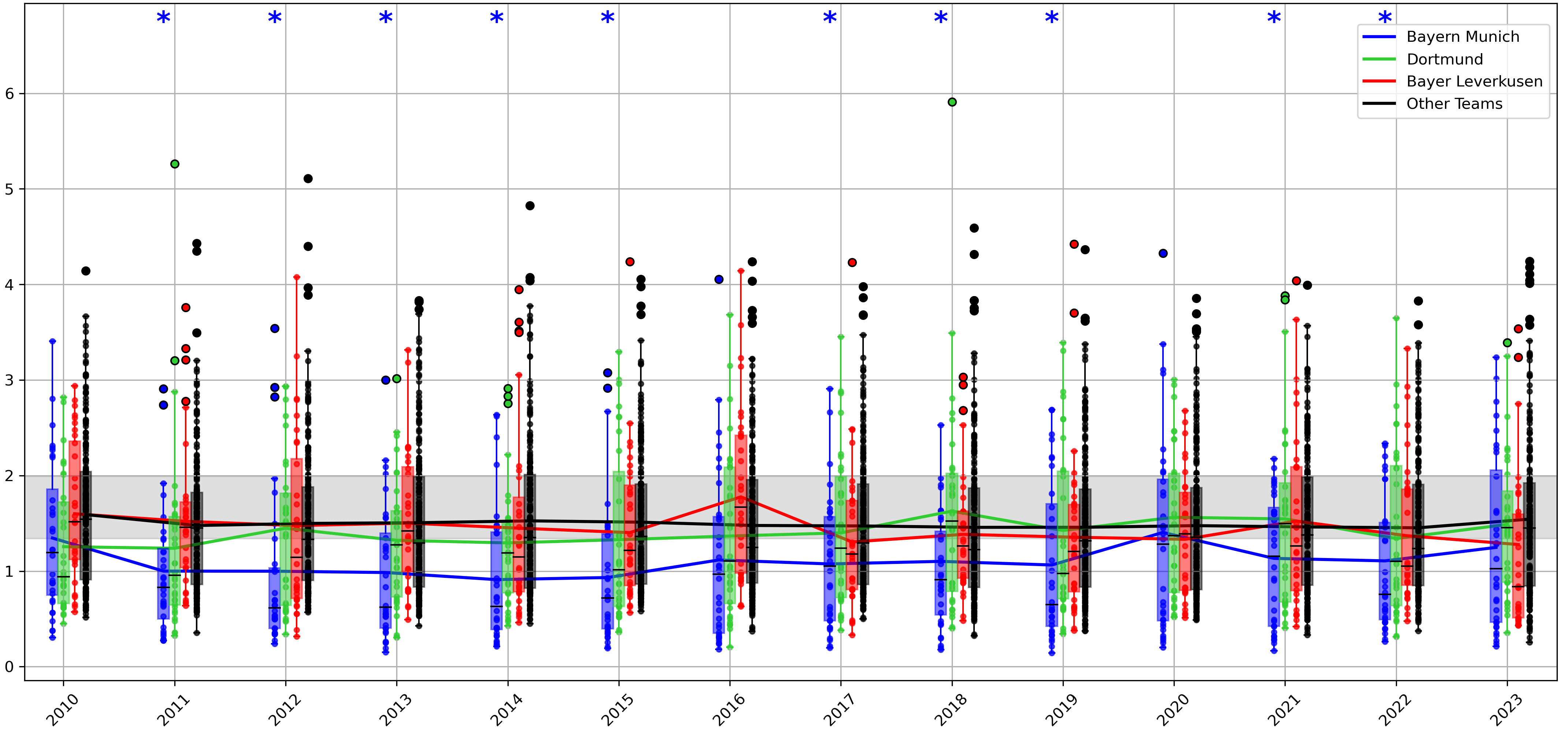}
        \caption{Surprise}
        \label{fig:surprise_gbl}
    \end{subfigure}
    \caption*{\footnotesize\textit{Notes}: This figure displays box plots for mean suspense (Figure \ref{fig:suspense_gbl}) and mean surprise (Figure \ref{fig:surprise_gbl}) per game across teams and seasons in the German Bundesliga. The grey horizontal band represents the benchmark range as defined in Section \ref{sec:sim_matches}. For suspense, the upper (lower) bound of the benchmark range corresponds to  mean suspense of an otherwise balanced match with home advantage and low (high) scoring rates of \( \lambda_H = 0.85 \) and \( \lambda_A = 0.5 \) (\( \lambda_H = 2.85, \ \lambda_A = 2.5 \)). For surprise, the upper (lower) bound of the benchmark range corresponds to mean surprise of an otherwise balanced match with home advantage and high (low) scoring rates of \( \lambda_H = 2.85 \) and \( \lambda_A = 2.5 \) (\( \lambda_H = 0.85, \ \lambda_A = 0.5 \)). * indicates a significant difference (\( p \) $< 0.05$) between the mean value at the team-season level and the lower bound of the benchmark range, based on a one-sided t-test.}
\end{figure}

Figure \ref{fig:suspense_surprise_gbl} presents box plots for the top three teams in the German Bundesliga—Bayern Munich, Borussia Dortmund, and Bayer Leverkusen—alongside the mean for all other teams. 

Figure \ref{fig:suspense_gbl} shows that mean suspense values for Bayern Munich matches are consistently the lowest in the Bundesliga and significantly below the benchmark range. Matches involving Dortmund exhibit the second-lowest suspense values, with many seasons also falling significantly below the lower bound. Thus, the Bundesliga is characterized by exceptionally low suspense in matches involving Bayern and Dortmund, while matches involving other teams remain on average closer to the lower benchmark bound associated with an otherwise balanced match with home advantage as defined in Section \ref{sec:sim_matches}. However, Panel A of Table \ref{tab:bundesliga_suspense_cond} in Appendix \ref{sec:team_seasons} also shows that particularly weak teams occasionally generated comparably low levels of suspense, such as Paderborn 07 (20 points in 2019/20), Schalke 04 (16 points in 2020/21), Greuther F\"urth (18 points in 2021/22), and Darmstadt 98 (17 points in 2023/24).

Interestingly, although the 2023/24 season saw a club other than Bayern win the title for the first time in 11 years, Bayern’s mean value of suspense remained lower than that of the eventual league winner, Bayer Leverkusen. Moreover, while mean surprise values commonly fall well within the benchmark range, they are significantly below this range for matches involving Bayern Munich in 10 of 14 seasons (see Figure \ref{fig:surprise_gbl}). 

The linear regression results in Table \ref{tab:time_trends_gbl} reveal an overall negative trend in both suspense and surprise in the Bundesliga. In the case of suspense, this trend is consistent with the greater number of team-seasons falling significantly below the benchmark range toward the end of our sample period (see Panel A of Table \ref{tab:bundesliga_suspense_cond} in Appendix \ref{sec:team_seasons}). However, the magnitudes of these effects appear modest (–0.7\% and –0.5\% per season, respectively) and do not appear to be driven by the top teams; if anything, the estimated trend for Dortmund is slightly positive.

\begin{table}[ht!]
    \centering
    \caption{German Bundesliga Time Trends}
    \label{tab:time_trends_gbl}
    \small
    \begin{tabular}{l c c c c}
        \toprule
        & \multicolumn{2}{c}{Ln(suspense)} & \multicolumn{2}{c}{Ln(surprise)} \\
        \cmidrule(lr){2-3} \cmidrule(lr){4-5}
        & (1) & (2) & (3) & (4) \\
        \midrule
        Season & $-$0.006*** & $-$0.007*** & $-$0.003 & $-$0.005** \\
               & (0.002)  & (0.002)  & (0.002)  & (0.002)  \\
        Bayern Munich × season &  & $-$0.003 &  & 0.012 \\
                               &  & (0.010)  &  & (0.009)  \\
        Borussia Dortmund × season &  & 0.012** &  & 0.016** \\
                                   &  & (0.006)  &  & (0.007)  \\
        Bayer Leverkusen × season &  & $-$0.009 &  & $-$0.012 \\
                                  &  & (0.006) &  & (0.007) \\
        \midrule
        Main effects &  & Yes &  & Yes \\
        Observations & 4,284 & 4,284 & 4,284 & 4,284 \\
        Adj-R\textsuperscript{2} & 0.002 & 0.101 & 0.000 & 0.058 \\
        \bottomrule
    \end{tabular}
    \caption*{\footnotesize\textit{Notes}: This table reports the results from OLS regressions with robust standard errors, clustered at the home team–away team level, shown in parentheses. Season is defined as a continuous variable. *** $p < 0.01$, ** $p < 0.05$, * $p < 0.1$.}
\end{table}

Overall, the Bundesliga’s title concentration, with Bayern winning the championship 11 times and Dortmund twice during our 14-year observation period, is clearly reflected in the observed suspense and surprise patterns. Importantly, however, while the levels of suspense and surprise have slightly decreased over time, this trend is not driven by either Bayern or Dortmund. 

\subsection{Suspense and Surprise Trends in the Spanish La Liga}

In Spain’s La Liga, Barcelona, Real Madrid, and Atlético Madrid have accumulated the most points during our observation period and are thus classified as the top teams. 

\begin{figure}[ht!]
    \centering
    \caption{Suspense and Surprise in the Spanish La Liga}
    \label{fig:suspense_surprise_sll}
    \begin{subfigure}{\textwidth}
        \centering
        \includegraphics[width=\linewidth,trim = 0mm 0mm 0mm 0mm, clip]{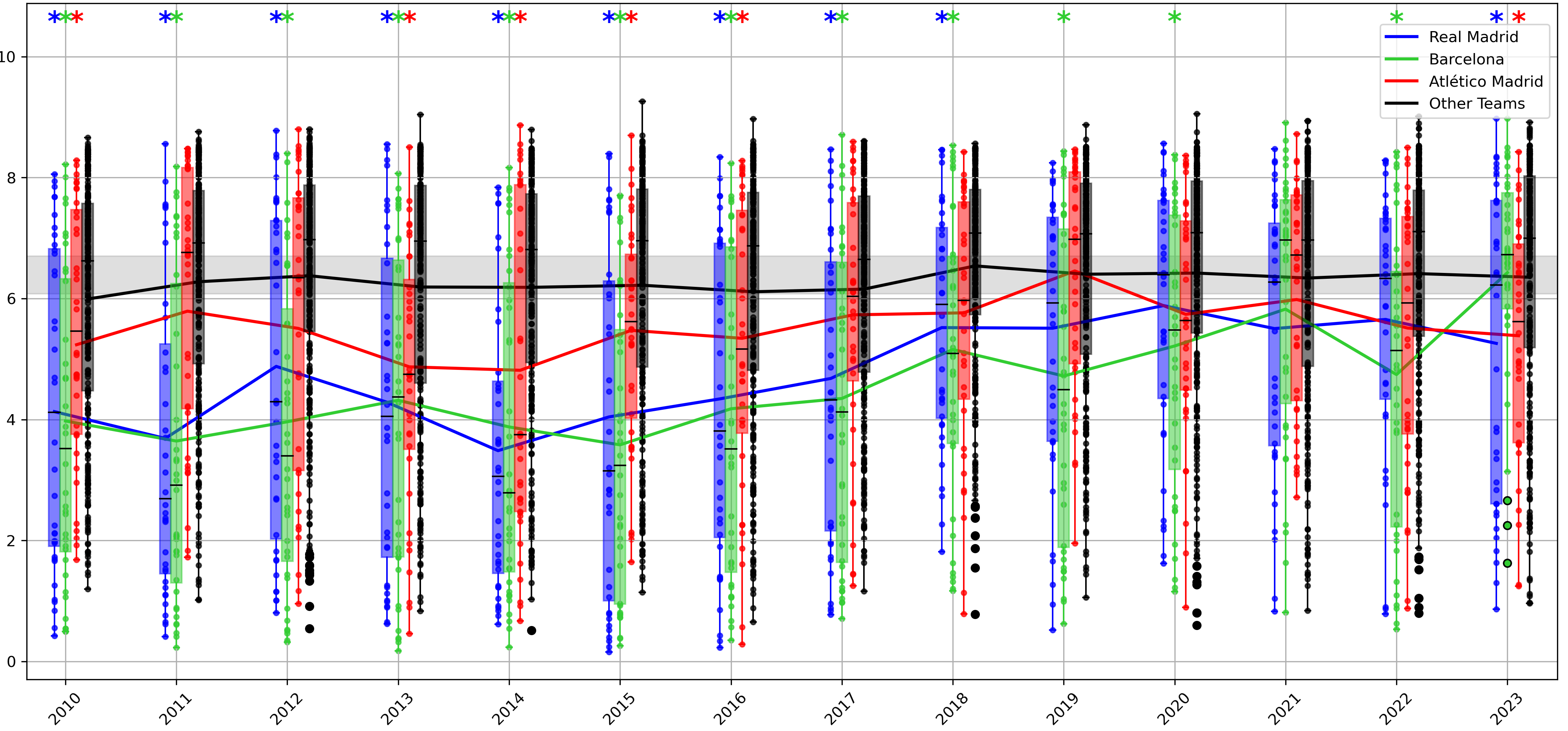}
        \caption{Suspense}
        \label{fig:suspense_sll}
    \end{subfigure}

    \begin{subfigure}{\textwidth}
        \centering
        \includegraphics[width=\textwidth,trim = 0mm 0mm 0mm 0mm, clip]{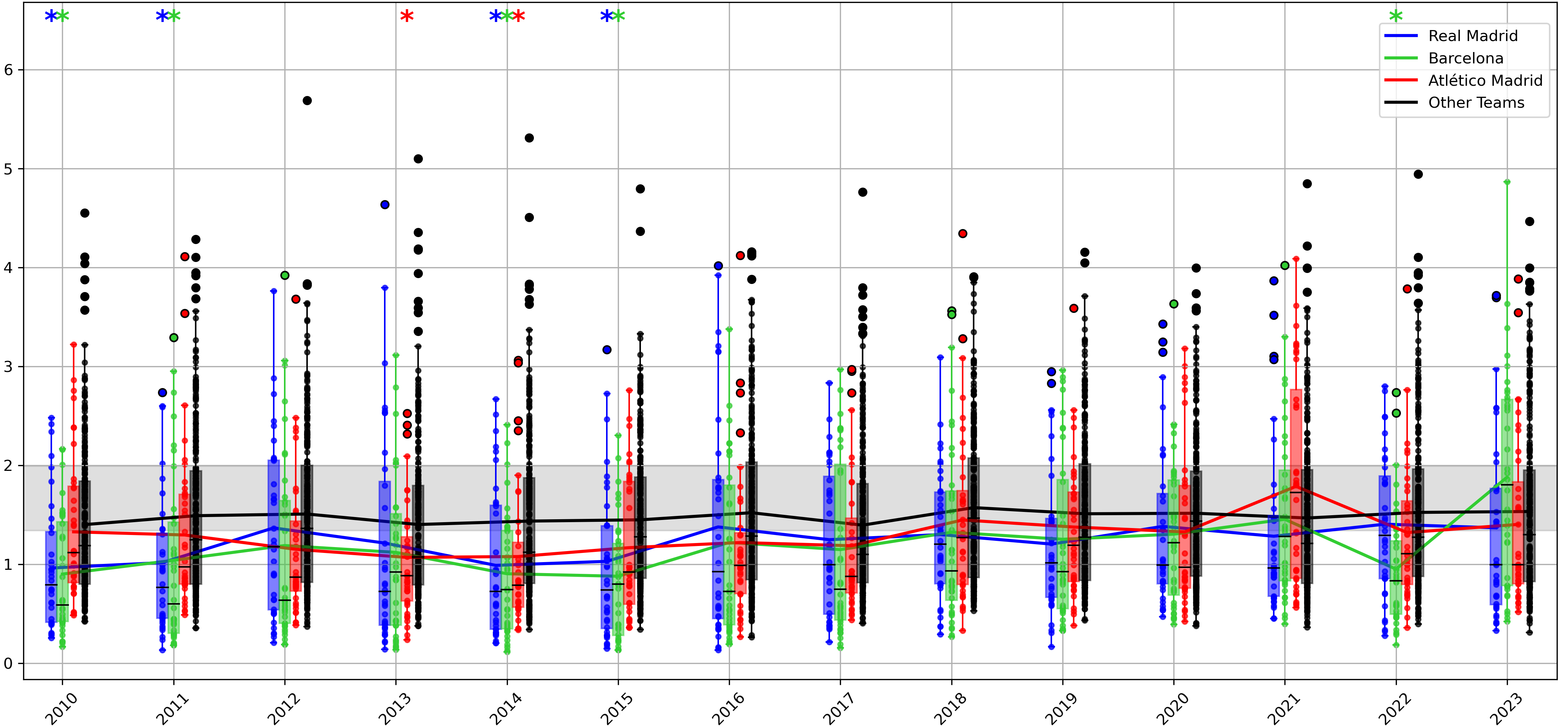}
        \caption{Surprise}
        \label{fig:surprise_sll}
    \end{subfigure}
    \caption*{\footnotesize\textit{Notes}: This figure displays box plots for mean suspense (Figure \ref{fig:suspense_sll}) and mean surprise  (Figure \ref{fig:surprise_sll}) per game across teams and seasons in the Spanish La Liga. The grey horizontal band represents the benchmark range as defined in Section \ref{sec:sim_matches}. For suspense, the upper (lower) bound of the benchmark range corresponds to  mean suspense of an otherwise balanced match with home advantage and low (high) scoring rates of \( \lambda_H = 0.85 \) and \( \lambda_A = 0.5 \) (\( \lambda_H = 2.85, \ \lambda_A = 2.5 \)). For surprise, the upper (lower) bound of the benchmark range corresponds to mean surprise of an otherwise balanced match with home advantage and high (low) scoring rates of \( \lambda_H = 2.85 \) and \( \lambda_A = 2.5 \) (\( \lambda_H = 0.85, \ \lambda_A = 0.5 \)). * indicates a significant difference (\( p \) $< 0.05$) between the mean value at the team-season level and the lower bound of the benchmark range, based on a one-sided t-test.}
\end{figure}

Figure \ref{fig:suspense_sll} shows that matches involving Barcelona or Real Madrid typically generate suspense values significantly below the benchmark range, although this pattern becomes less pronounced in the second half of the sample period. 

In the 2023/24 season, Barcelona’s suspense values increased sharply and fell within the benchmark range; however, in 2021/22 they were already not significantly different from the lower bound of the benchmark range. For Atlético Madrid, suspense values tend to be higher compared to the other top clubs, but still fall significantly below the lower benchmark bound in most of the earlier years. The surprise box plots for La Liga, shown in Figure \ref{fig:surprise_sll}, are broadly consistent with previous observations. While Barcelona’s surprise values fall significantly below the benchmark range on several occasions, the values for all other teams, including Real Madrid and Atlético Madrid, mostly lie within the benchmark range. In the 2023/24 season, however, Barcelona’s surprise values increased markedly, surpassing even the mean value of all other teams.

Remarkably, and in contrast to the Bundesliga, a greater number of team-seasons fall significantly below the benchmark range at the beginning rather than at the end of our sample period (see Panels A and B of Table \ref{tab:laliga_suspense_cond} in Appendix \ref{sec:team_seasons}). Consistent with this pattern, visual inspection of Figure \ref{fig:suspense_sll} suggests that mean suspense slightly increases over time. 

This is confirmed by Table \ref{tab:time_trends_sll}, which indicates a mild overall increase in suspense for non-top teams of approximately 0.5\% per season (see Column 2). This positive trend is much more pronounced in matches involving Real Madrid or Barcelona, which exhibit additional increases in suspense of 4.0\% and 4.9\% per season, respectively, relative to the baseline trend. A similar upward pattern is observed for surprise. As shown in Column (4) of Table \ref{tab:time_trends_sll}, matches involving Real Madrid and Barcelona exhibit additional increases in surprise of 2.3\% and 3.8\% per season, respectively, relative to the baseline increase of 0.6\% per season.

\begin{table}[ht!]
    \centering
    \caption{Spanish La Liga Time Trends}
    \label{tab:time_trends_sll}
    \small
    \begin{tabular}{l c c c c}
        \toprule
        & \multicolumn{2}{c}{Ln(suspense)} & \multicolumn{2}{c}{Ln(surprise)} \\
        \cmidrule(lr){2-3} \cmidrule(lr){4-5}
        & (1) & (2) & (3) & (4) \\
        \midrule
        Season & 0.014*** & 0.005*** & 0.013*** & 0.006*** \\
               & (0.002)  & (0.002)  & (0.002)  & (0.002)  \\
        Real Madrid × season &  & 0.040*** &  & 0.023*** \\
                         &  & (0.009)  &  & (0.008)  \\
        Barcelona × season &  & 0.049*** &  & 0.038*** \\
                          &  & (0.010)   &  & (0.009)   \\
        Atlético Madrid × season &  & 0.004 &  & 0.010 \\
                                 &  & (0.007) &  & (0.008) \\
        \midrule
        Main effects &  & Yes &   & Yes \\
        Observations & 5,320 & 5,320 & 5,320 & 5,320 \\
        Adj-R\textsuperscript{2} & 0.010 & 0.122 & 0.008 & 0.061 \\
        \bottomrule
    \end{tabular}
    \caption*{\footnotesize\textit{Notes}: This table reports the results from OLS regressions with robust standard errors, clustered at the home team–away team level, shown in parentheses. Season is defined as a continuous variable. *** $p < 0.01$.}
\end{table}

Overall, suspense values, and to a lesser extent surprise values, were substantially below the benchmark range at the start of the sample period for La Liga matches, particularly those involving top teams. However, both measures have steadily increased over time, with the most pronounced growth observed for Real Madrid and Barcelona.

\subsection{Suspense and Surprise Trends in the Italian Serie A}

\begin{figure}[ht!]
    \centering
    \caption{Suspense and Surprise in the Italian Serie A}
    \label{fig:suspense_surprise_isa}
    \begin{subfigure}{\textwidth}
        \centering
        \includegraphics[width=\linewidth,trim = 0mm 0mm 0mm 0mm, clip]{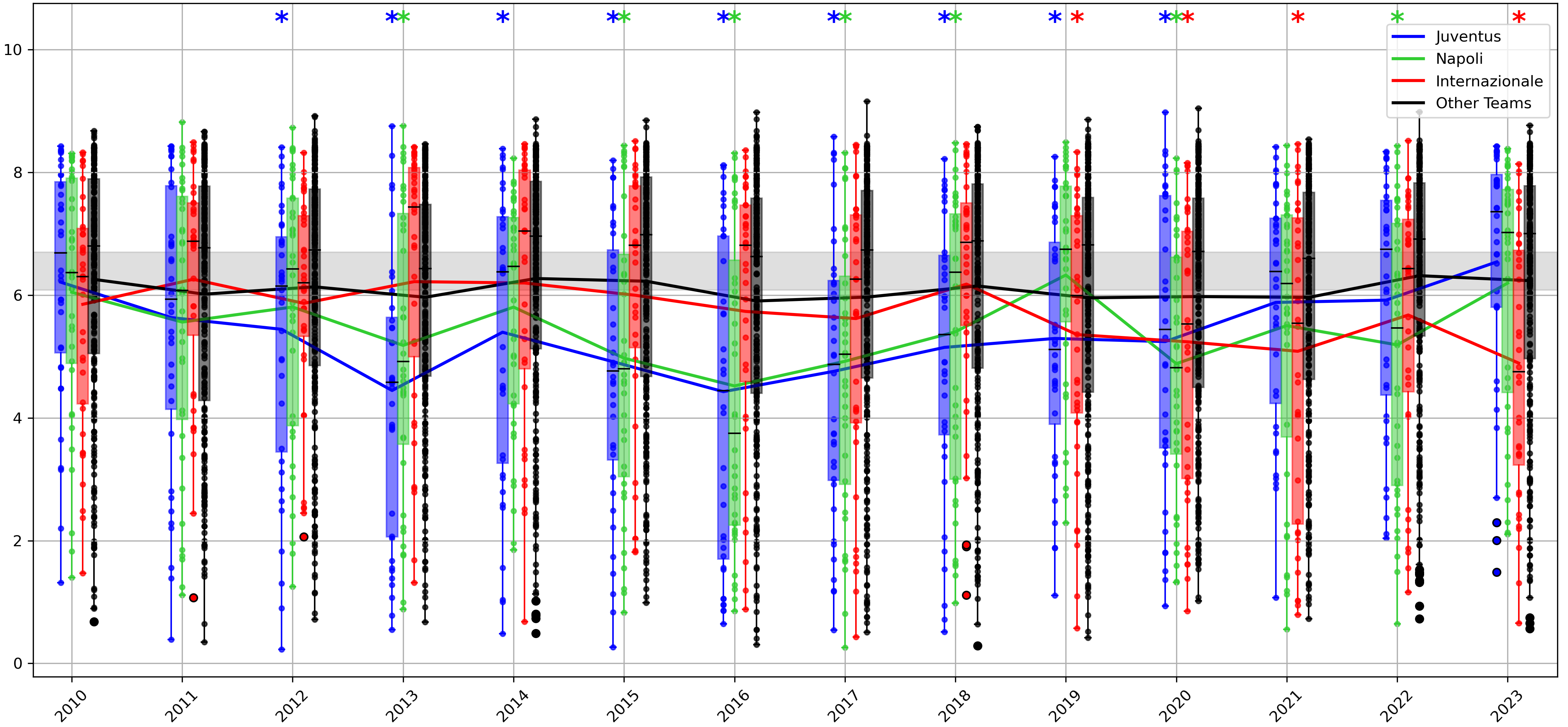}
        \caption{Suspense}
        \label{fig:suspense_isa}
    \end{subfigure}

    \begin{subfigure}{\textwidth}
        \centering
        \includegraphics[width=\textwidth,trim = 0mm 0mm 0mm 0mm, clip]{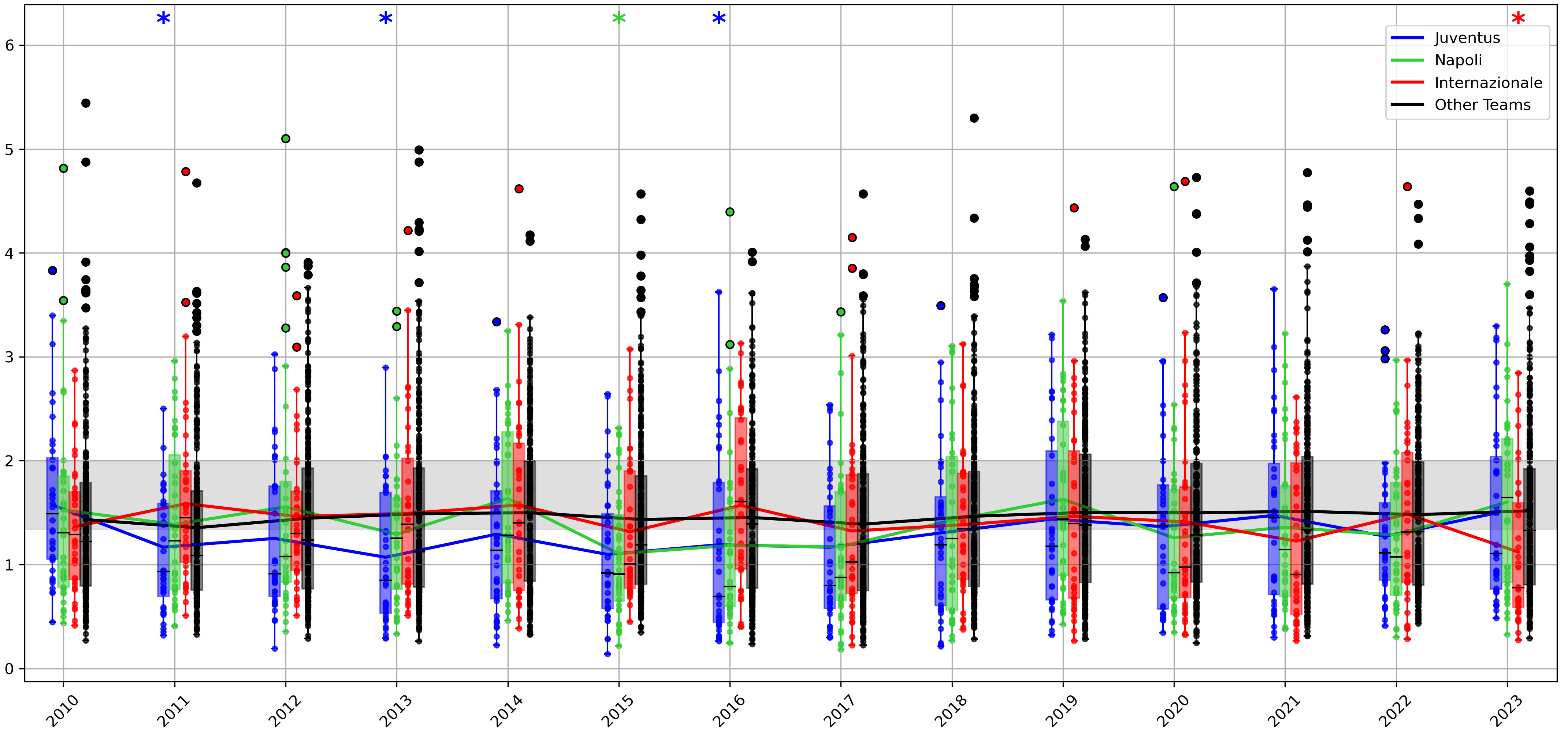}
        \caption{Surprise}
        \label{fig:surprise_isa}
    \end{subfigure}
    \caption*{\footnotesize\textit{Notes}: This figure displays box plots for mean suspense (Figure \ref{fig:suspense_isa}) and mean surprise \ref{fig:surprise_isa}) per game across teams and seasons in the Italian Serie A. The grey horizontal band represents the benchmark range as defined in Section \ref{sec:sim_matches}. For suspense, the upper (lower) bound of the benchmark range corresponds to  mean suspense of an otherwise balanced match with home advantage and low (high) scoring rates of \( \lambda_H = 0.85 \) and \( \lambda_A = 0.5 \) (\( \lambda_H = 2.85, \ \lambda_A = 2.5 \)). For surprise, the upper (lower) bound of the benchmark range corresponds to mean surprise of an otherwise balanced match with home advantage and high (low) scoring rates of \( \lambda_H = 2.85 \) and \( \lambda_A = 2.5 \) (\( \lambda_H = 0.85, \ \lambda_A = 0.5 \)). * indicates a significant difference (\( p \) $< 0.05$) between the mean value at the team-season level and the lower bound of the benchmark range, based on a one-sided t-test.}
\end{figure}

Figure \ref{fig:suspense_surprise_isa} presents box plots for the top teams, Juventus, Napoli, and Internazionale, alongside the mean values of all other teams in the Italian Serie A.

Figure \ref{fig:suspense_isa} shows that suspense values in the 2010/11 season were initially clustered closely together, just around the lower bound of the benchmark range. Over time, however, particularly in matches involving Juventus, suspense values regularly fell below the benchmark. Toward the end of the sample period, suspense values appear to converge again, with most teams falling within the benchmark range, except for Internazionale, which won, for instance, the 2023/24 season by a margin of 19 points. Interestingly, in 2023/24, Juventus recorded the highest mean of suspense, which was even higher than the mean value of all other teams and well within the benchmark range. With regard to surprise, Figure \ref{fig:surprise_isa} shows that values most often remained within the benchmark range. However, several lower-ranked teams again recorded isolated team-seasons with low suspense and/or surprise values, including Pescara (2012/13 and 2016/17, with 22 and 18 points, respectively), Benevento (2017/18, with 21 points), and Hellas Verona (2017/18, with 25 points). For a full list of team-seasons with mean suspense and/or surprise values that fall significantly below the benchmark range, see Panels A and B of Table \ref{tab:seriea_suspense_cond} in Appendix \ref{sec:team_seasons}.

The linear regression results in Table~\ref{tab:time_trends_isa} do not indicate any notable overall trends in Serie A. However, matches involving Juventus exhibit an additional increase in suspense of approximately 1.4\% per season relative to the baseline trend. By contrast, matches involving Internazionale show additional declines in suspense and surprise of approximately 2.1\% and 2.4\% per season, respectively.

\begin{table}[ht!]
    \centering
    \caption{Italian Serie A Time Trends}
    \label{tab:time_trends_isa}
    \small
    \begin{tabular}{l c c c c}
        \toprule
        & \multicolumn{2}{c}{Ln(suspense)} & \multicolumn{2}{c}{Ln(surprise)} \\
        \cmidrule(lr){2-3} \cmidrule(lr){4-5}
        & (1) & (2) & (3) & (4) \\
        \midrule
        Season & $-$0.002 & $-$0.001 & 0.002 & 0.004* \\
               & (0.002)  & (0.002)  & (0.002)  & (0.002)  \\
        Juventus × season &  & 0.014** &  & 0.006 \\
                        &  & (0.007)  &  & (0.007)  \\
        Napoli × season &  & $-$0.001 &  & $-$0.005 \\
                                &  & (0.006)  &  & (0.007)  \\
        Internazionale × season &  & $-$0.021*** &  & $-$0.024*** \\
                      &  & (0.006) &  & (0.007) \\
        \midrule
        Main effects & & Yes &   &  Yes\\
        Observations &  5,320 &  5,320 &  5,320 &  5,320 \\
        Adj-R\textsuperscript{2} & 0.000 & 0.017 & 0.000 & 0.010\\
        \bottomrule
    \end{tabular}
    \caption*{\footnotesize\textit{Notes}: This table reports the results from OLS regressions with robust standard errors, clustered at the home team–away team level, shown in parentheses. Season is defined as a continuous variable. *** $p < 0.01$, ** $p < 0.05$, * $p < 0.1$.}
\end{table}

\subsection{Suspense and Surprise Trends in the French Ligue 1}

\begin{figure}[ht!]
    \centering
    \caption{Suspense and Surprise in the French Ligue 1}
    \label{fig:suspense_surprise_fl1}
    \begin{subfigure}{\textwidth}
        \centering
        \includegraphics[width=\linewidth,trim = 0mm 0mm 0mm 0mm, clip]{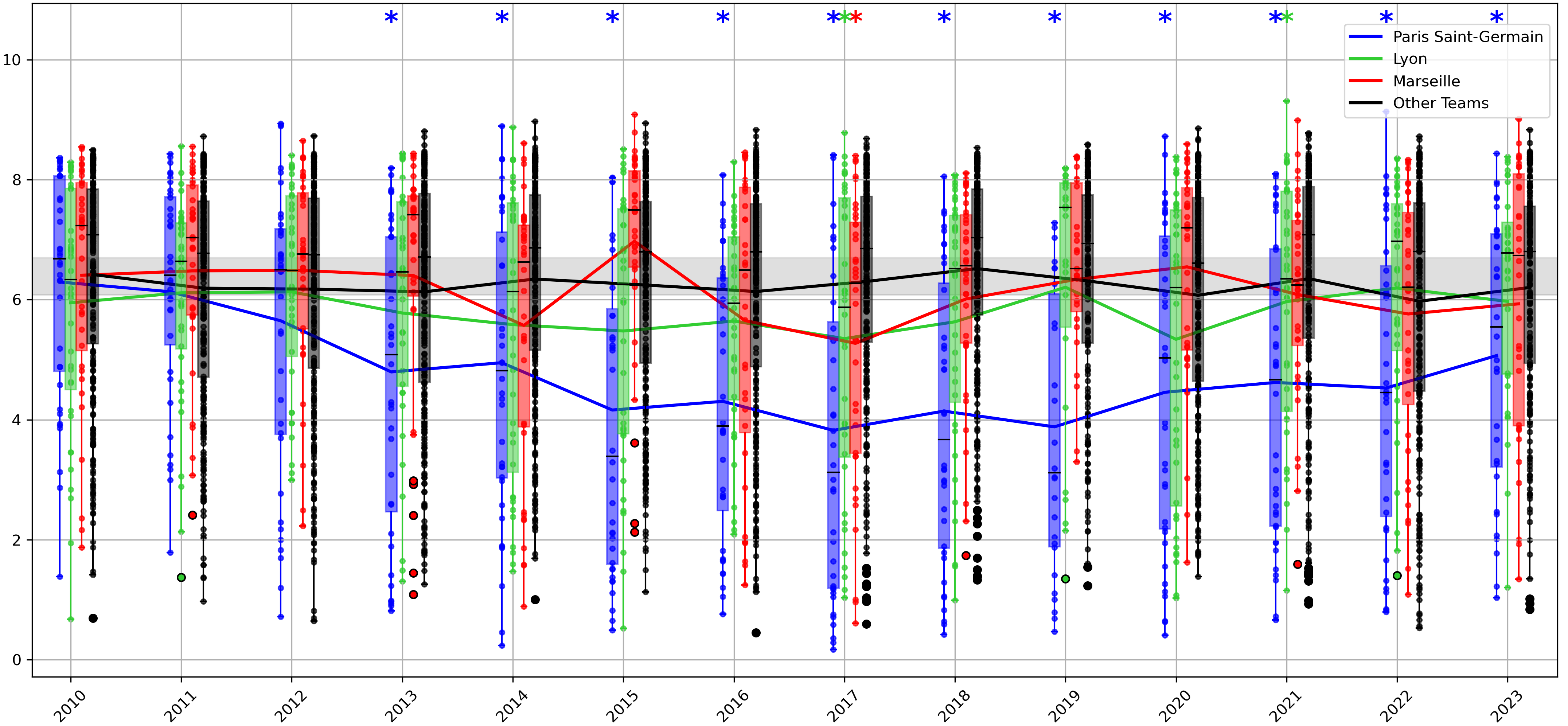}
        \caption{Suspense}
        \label{fig:suspense_fl1}
    \end{subfigure}

    \begin{subfigure}{\textwidth}
        \centering
        \includegraphics[width=\textwidth,trim = 0mm 0mm 0mm 0mm, clip]{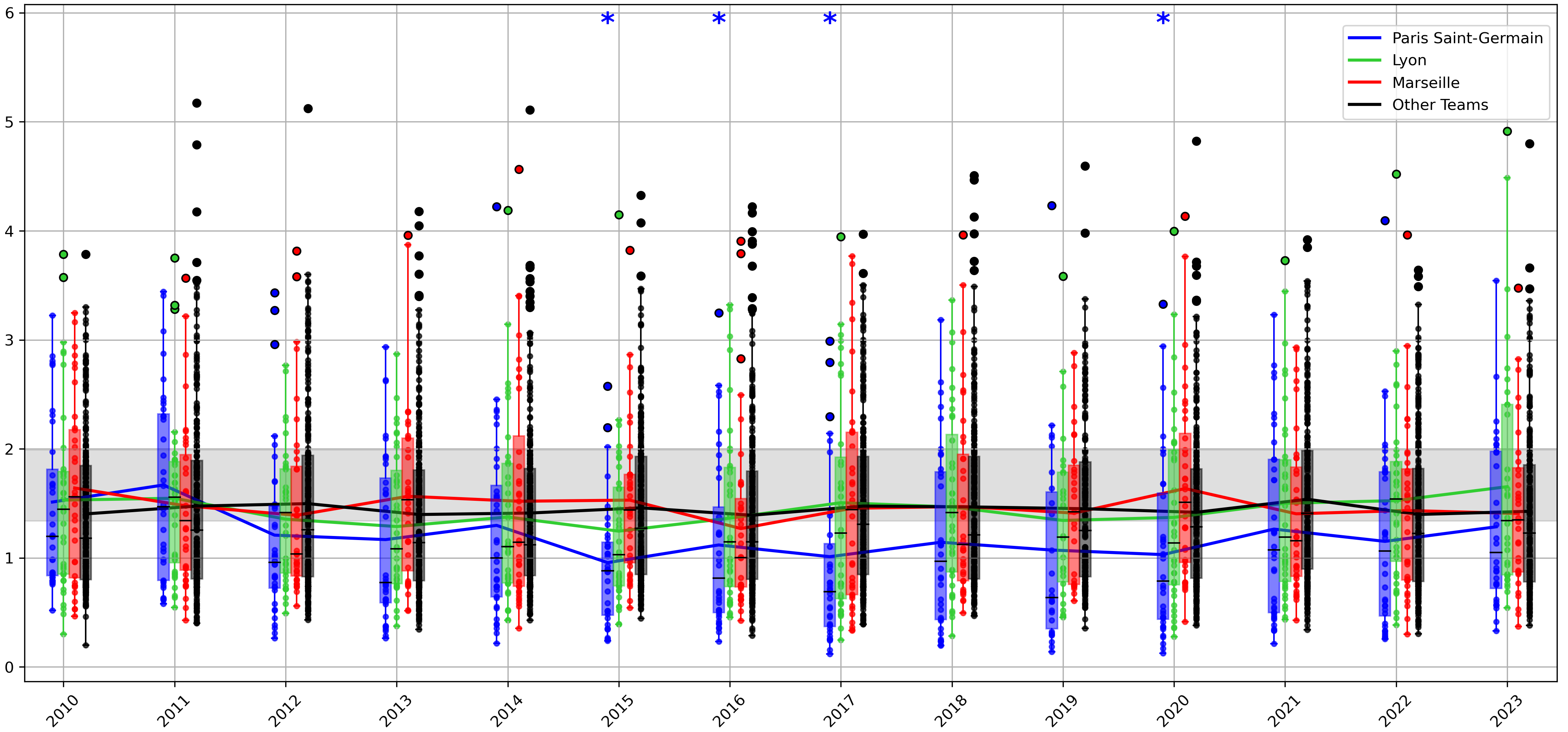}
        \caption{Surprise}
        \label{fig:surprise_fl1}
    \end{subfigure}
    \caption*{\footnotesize\textit{Notes}: This figure displays box plots for mean suspense (Figure \ref{fig:suspense_fl1}) and mean surprise (Figure \ref{fig:surprise_fl1}) per game across teams and seasons in the French Ligue 1. The grey horizontal band represents the benchmark range as defined in Section \ref{sec:sim_matches}. For suspense, the upper (lower) bound of the benchmark range corresponds to  mean suspense of an otherwise balanced match with home advantage and low (high) scoring rates of \( \lambda_H = 0.85 \) and \( \lambda_A = 0.5 \) (\( \lambda_H = 2.85, \ \lambda_A = 2.5 \)). For surprise, the upper (lower) bound of the benchmark range corresponds to mean surprise of an otherwise balanced match with home advantage and high (low) scoring rates of \( \lambda_H = 2.85 \) and \( \lambda_A = 2.5 \) (\( \lambda_H = 0.85, \ \lambda_A = 0.5 \)). * indicates a significant difference (\( p \) $< 0.05$) between the mean value at the team-season level and the lower bound of the benchmark range, based on a one-sided t-test.}
\end{figure}

Figure \ref{fig:suspense_surprise_fl1} shows box plots for the top teams Paris Saint-Germain, Lyon, and Marseille, together with the mean values for all other teams in French Ligue 1.

In the French Ligue 1, only suspense values in matches involving Paris Saint-Germain frequently fall significantly below the lower bound of the benchmark range, as also confirmed by Table \ref{tab:ligue1_suspense_cond} in Appendix \ref{sec:team_seasons}. Interestingly, the suspense and surprise values of all teams were initially clustered closely together during the first three seasons in our sample (2010/11 to 2012/13), before the mean values for Paris Saint-Germain began to decline substantially.

The linear regressions in Table \ref{tab:time_trends_fl1} confirm these visual patterns. While Column (1) indicates a mild negative overall trend in suspense, Column (2) shows that this trend is driven mainly by matches involving Paris Saint-Germain. These matches exhibit an additional decline in suspense of approximately 2.9\% per season relative to the baseline trend. A similar pattern emerges for surprise: as shown in Column (4), matches involving Paris Saint-Germain exhibit an additional decline of approximately 2.5\% per season relative to the baseline trend. By contrast, we do not observe systematic changes for matches involving Lyon, Marseille, or the remaining teams in Ligue 1.

\begin{table}[ht!]
    \centering
    \caption{French Ligue 1 Time Trends}
    \label{tab:time_trends_fl1}
    \small
    \begin{tabular}{l c c c c}
        \toprule
        & \multicolumn{2}{c}{Ln(suspense)} & \multicolumn{2}{c}{Ln(surprise)} \\
        \cmidrule(lr){2-3} \cmidrule(lr){4-5}
        & (1) & (2) & (3) & (4) \\
        \midrule
        Season & $-$0.006*** & $-$0.002 & $-$0.002 & 0.001 \\
               & (0.002)  & (0.002)  & (0.002)  & (0.002)  \\
        Paris Saint-Germain × season &  & $-$0.029*** &  & $-$0.025*** \\
                                     &  & (0.007)   &  & (0.009)   \\
        Lyon × season &  & 0.002 &  & 0.002 \\
                      &  & (0.005) &  & (0.006) \\
        Marseille × season &  & $-$0.004 &  & $-$0.005 \\
                           &  & (0.004) &  & (0.007) \\
        \midrule
        Main effects & &  Yes &  &  Yes\\
        Observations & 5,145 & 5,145 & 5,145 & 5,145 \\
        Adj-R\textsuperscript{2} & 0.002 & 0.068 & 0.000 & 0.027 \\
        \bottomrule
    \end{tabular}
    \caption*{\footnotesize\textit{Notes}: This table reports the results from OLS regressions with robust standard errors, clustered at the home team–away team level, shown in parentheses. Season is defined as a continuous variable. *** $p < 0.01$.}
\end{table}

\subsection{Summary of Empirical Findings}

Table \ref{tab:summary_level_trend} summarizes our empirical findings. Panel A reports the number of team-seasons in which mean suspense or mean surprise falls below the benchmark ranges defined in Section \ref{sec:sim_matches}, while Panel B reports the corresponding time trends derived from the regression analyses presented above.

\begin{table}[ht]
    \singlespacing
    \centering 
    \caption{Summary Table for Levels and Time Trends}
    \label{tab:summary_level_trend}
    \begin{tabular}{l c c c c c c}
        \toprule
        \multicolumn{7}{l}{\textit{Panel A: Number of team-seasons $<$ BM}} \\
        \midrule 
        & \multicolumn{3}{c}{Suspense} 
        & \multicolumn{3}{c}{Surprise} \\
        \cmidrule(lr){2-4} \cmidrule(lr){5-7}
        & All & Top 3 & Other & All & Top 3 & Other \\
        \midrule
        Premier League & 63 & 26 & 37 (12) & 13 & 7 & 6 (4) \\

        Bundesliga     & 39 & 24 & 15 (6) & 12 & 10 & 2 (0) \\

        La Liga        & 48 & 28 & 20 (6) & 15 & 11 & 4 (1) \\

        Serie A        & 49 & 20 & 29 (9) & 11 & 5 & 6 (2) \\

        Ligue 1        & 20 & 14 & 6 (3) &  9 & 4 & 5 (3) \\
        \midrule
        Total          & 219 & 112 & 107 (36) & 60 & 37 & 23 (10) \\
        \midrule
\multicolumn{7}{l}{\textit{Panel B: Time trends}} \\
       \midrule
        & \multicolumn{3}{c}{Suspense} 
        & \multicolumn{3}{c}{Surprise} \\
        \cmidrule(lr){2-4} \cmidrule(lr){5-7}
        & All & Top 3 & Other & All & Top 3 & Other \\
        \midrule
        Premier League & \downarrowred & \downarrowred\ -- -- & -- & -- & \downarrowred\ -- -- & -- \\
   
        Bundesliga     & \downarrowred & -- \uparrowgreen\ -- & \downarrowred & -- & -- \uparrowgreen\ -- & \downarrowred \\

        La Liga        & \uparrowgreen & \uparrowgreen\ \uparrowgreen\ -- & \uparrowgreen & \uparrowgreen & \uparrowgreen\ \uparrowgreen\ -- & \uparrowgreen \\
        Serie A        & -- & \uparrowgreen\ -- \downarrowred & -- & -- & -- -- \downarrowred & \uparrowgreen \\
        Ligue 1        & \downarrowred & \downarrowred\ -- -- & -- & -- & \downarrowred\ -- -- & -- \\
        \bottomrule
    \end{tabular}
\caption*{\footnotesize\textit{Notes}: ``$<$ BM" refers to the result of a one-sided \textit{t}-test assessing whether the mean for a team in a season is below the benchmark (BM) of 6.08 for suspense and 1.34 for surprise. ``All" refers to all teams in the league in a given season. ``Top 3" refers to those teams that accumulated the most points during our observation period from 2010/11 to 2023/24. ``Other" refers to all non-top 3 teams in a given season. The top 3 teams are the following (ordered by the 1\textsuperscript{st}, 2\textsuperscript{nd}, and 3\textsuperscript{rd} best performing teams): Premier League: Manchester City, Liverpool, and Arsenal; Bundesliga: Bayern Munich, Borussia Dortmund, and Bayer Leverkusen; La Liga: Barcelona, Real Madrid, and Atl\'etico Madrid; Serie A: Juventus, Napoli, and Internazionale; Ligue 1: Paris Saint-Germain, Lyon, and Marseille. Values in parentheses indicate the number of observations in the ``Other'' category involving teams relegated in the corresponding season. Time trends are derived from the regressions (see Sections 4.2-4.6) and coded as $\uparrow$ (upward), $\downarrow$ (downward), and -- (no trend). For top 3 teams, the trends are presented by rank, i.e. (from left to right), the 1\textsuperscript{st} (2\textsuperscript{nd}) [3\textsuperscript{rd}] trend refers to the 1\textsuperscript{st} (2\textsuperscript{nd}) [3\textsuperscript{rd}] best performing team during our observation window.}
\end{table}

While the leagues differ little in the number of team-seasons falling below the benchmark for surprise, more pronounced differences emerge for suspense. For example, over our observation window, the Premier League accumulated more than three times as many such team-seasons as Ligue 1 (63 vs. 20). This difference is driven by both top teams (26 vs. 14) and other teams (37 vs. 6). Remarkably, the corresponding number for other teams in the Premier League (37) is also considerably higher than in the other three leagues: 15 in the Bundesliga, 20 in La Liga, and 29 in Serie A. Moreover, across all leagues, these counts for other teams are driven by both particularly strong teams outside the top 3 and particularly weak teams in a given season. This reflects the general nature of our measures: suspense and surprise are low both when the favorite wins comfortably and when the underdog loses clearly.

Our trend analyses reveal three distinct patterns across and within the leagues. First, we identify leagues in which both mean suspense and surprise decline markedly in matches involving a single top team, while matches featuring other teams in the same league remain relatively stable. This pattern applies to Manchester City in the Premier League and Paris Saint-Germain in Ligue 1, and to a lesser extent to Internazionale in Serie A. Second, we observe a modest overall decline in both mean suspense and surprise in the German Bundesliga, which is not driven by top teams. Third, we identify a marked increase in mean suspense and surprise over time in Spain’s La Liga, a trend that appears to be accentuated by matches involving Real Madrid and Barcelona.

\section{Discussion and Conclusion}\label{sec:dis_concl}

We present match-level suspense and surprise as measures that capture the entertainment utility generated by competitive balance and outcome uncertainty for sports spectators, as they account for how outcome uncertainty unfolds throughout the course of a match. These measures seem promising because they offer and accurate understanding of both the current state and the dynamics of match attractiveness associated with competitive balance across clubs and seasons. As such, our micro analysis complements exiting macro analyses which regularly focus on much more aggregated measures of competitive balance.

Through simulation analysis, we identify a trade-off between suspense and surprise even in perfectly balanced matches: suspense tends to be higher at lower scoring rates, while surprise increases with higher scoring rates. Building on this insight, we establish benchmark ranges for both metrics based on otherwise perfectly balanced matches with home advantage and varying scoring rates. Our analysis of the top five men's European football leagues indicates that mean suspense is commonly below the benchmark range, particularly for games involving top teams. In contrast, matches involving non-top teams tend to show mean suspense values that are close to the benchmark’s lower bound or fall within the benchmark range. Meanwhile, mean surprise values mostly remain within the benchmark range.

This empirical pattern can be interpreted in light of further simulation results showing that suspense is substantially more sensitive to team-strength imbalance than surprise when holding the average scoring-rate level constant. Since matches involving top teams are likely to involve greater team-strength imbalances, such imbalances may reduce suspense more strongly than surprise. This provides a plausible explanation for why mean match suspense is often below the benchmark range, whereas mean match surprise more often remains within the benchmark range. However, this pattern should not be interpreted as implying that surprise is inherently more important to consumers than suspense. Rather, it indicates that suspense and surprise respond differently to scoring rates and competitive imbalance.

In contrast to other studies that report a general decline in competitive balance in European football over time \citep[e.g.,][]{AvilaCanoTrigueroRuiz2023, Basini2023, Pawlowski2010, Poli2018}, we do not find such overall decline when using suspense and surprise. Rather we observe considerable differences across leagues including downward, upward, or no trends at all. This is in line with \citet{Csato2025} who did not find any evidence for declining competitive balance in the UEFA Champions League group stage using novel indicators of competitive balance. Moreover, some trends we observe for the whole league can be fully attributed to matches involving just one or two clubs within these leagues. However, it is important to note that lower levels of mean suspense and surprise in matches featuring top teams do not currently appear to reduce overall sporting interest. The negative impact may be offset by positive influences such as star appeal or the quality of play \citep[e.g.,][]{Buraimo2022}.

Overall, our paper provides a practical framework for assessing the state and evolution of suspense and surprise in European football. This approach not only advances the theoretical understanding of how competitive balance and outcome uncertainty translate into demand for sports, but also offers actionable insights. In particular, we outline three key practical implications. First, policymakers and league organizers should consider the trade-offs between suspense and surprise when designing interventions aimed at increasing the attractiveness of their competition. For example, efforts to equalize team strengths may increase suspense but reduce surprise, depending on the absolute level of scoring rates. Second, our findings highlight the importance of examining not only structural differences across leagues but also club-level variation over time. Mean values of these measures can differ substantially even within a single league, suggesting that league-wide trends may obscure important patterns at the level of individual clubs. Third, although few top clubs financially dominate their leagues, this dominance is only partially reflected in lower suspense and surprise values, which continue to exhibit substantial match-level variation. One possible explanation lies in the low-scoring nature of football, where chance plays a significant role \citep[e.g.,][]{Brechot2020}. As a result, the current levels of suspense and surprise in men’s football may not suggest an urgent need for regulatory intervention. Nevertheless, continued monitoring may be warranted given (i) downward trends in mean suspense and surprise for certain teams during our observation window from 2010/11 to 2023/24, for example, in matches involving Manchester City, Paris Saint-Germain, and Internazionale, as well as persistently low levels in matches involving Bayern Munich, and (ii) the possibility that these trends may change as additional seasons are taken into account \citep{Csato2025}.

We conclude by acknowledging several limitations of our study and highlighting promising directions for future research. One limitation lies in our focus on match-level suspense and surprise, which captures only short-term, match-specific dynamics. Extending this perspective to the medium term, for example, by calculating suspense and surprise with respect to beliefs about league outcomes or title races, could provide valuable insights into how season-long narratives shape sporting demand. Another limitation is that the benchmark ranges we develop are purely theory-based and have not been validated directly against consumer demand. Future research could therefore assess these benchmark ranges using behavioral data from fans to identify potential “sweet spots” of match-level suspense and surprise from the consumer perspective.

Methodologically, our in-play probability model relies on betting odds, which may be subject to various biases \citep[e.g.,][]{Feddersen2017}. However, a recent paper by \cite{Winkelmann2024} shows that, although informational inefficiencies in betting odds may occur in individual seasons, they are neither persistent nor systematic across the top five European football leagues. We therefore do not believe that our key results are driven by systematic biases in betting odds. Moreover, our in-play probability model incorporates only goals and red cards, omitting other potentially influential in-game events. Incorporating data on missed penalties, other scoring opportunities, corner kicks, possession changes and possession streaks, or defensive interventions could lead to more refined measures of suspense and surprise. In this context, recent advancements in modeling expected possession values \citep[e.g.,][]{Fernandez2021} represent a promising avenue for further development. A further limitation is that our analysis adopts a neutral perspective, without accounting for the emotional attachment that fans (or ``haters") may have toward different clubs. These emotional factors likely shape the perception of suspense and surprise, making fan loyalty an important consideration for future research, particularly in the context of league governance \citep[e.g.,][]{Pawlowski2024}. 

Finally, extending the analysis beyond men’s football appears to be an important avenue for future research. One natural step would be to examine women’s football. This could also be revealing from a methodological perspective, as recent findings question whether women’s football can be adequately modeled using univariate Poisson distributions \citep{Michels2025}. Another promising extension would be to apply the analysis to sports other than football. The concepts of suspense and surprise are broadly applicable across sports, including non-invasive sports. For high-scoring sports such as basketball, for example, it may also be useful to examine complementary micro-level measures, such as the number of lead changes during a game Overall, such comparative studies could offer valuable insights into how suspense and surprise function across diverse sporting environments, thereby enriching our broader understanding of professional sport as an entertainment product.

\bibliographystyle{apacite}
\bibliography{reference_list_sus_sur}

\clearpage
\appendix
\renewcommand{\thefigure}{\thesection.\arabic{figure}}
\renewcommand{\thetable}{\thesection.\arabic{table}}
\makeatletter
\@addtoreset{figure}{section}
\@addtoreset{table}{section}
\makeatother

\begin{center}
{\LARGE\bfseries Online Appendix}

\vspace{0.5em}

{\LARGE Suspense and Surprise in European Football}

\vspace{0.75em}

{\large Raphael Flepp \qquad Tim Pawlowski \qquad Travis Richardson}
\end{center}

\addcontentsline{toc}{section}{Appendix}

\vspace{1em}

\section{TV Demand Analysis}\label{sec:tvdemand}

We begin our validity check of suspense and surprise by merging our match-level measures for the Premier League with the original dataset used by \citet{Buraimo2022}. This dataset includes all matches broadcast in the UK market on Sky Sports and BT Sport between the second half of the 2013/14 season and the end of the 2018/19 season (\( n=790 \)). Relative to all other Premier League matches played between the 2013/14 and 2018/19 seasons, suspense values in this subsample are significantly lower (\textit{t}(2,278) = 3.26, \(p\) $<$ 0.001), whereas surprise values do not differ significantly (\textit{t}(2,278) = 0.33, \(p\) = 0.74).

The match audience size is defined as the average per-minute audience size, measured from kick-off to 110 minutes later, accounting for half-time (15 minutes) and added extra time (5 minutes). In addition, the dataset includes match-level measures of player quality derived from an expected points plus-minus rating model, match significance measures based on match forecasting models, and a commonly used measure of outcome uncertainty based on the absolute difference in pre-match win probabilities for the two teams \citep[e.g.,][]{Caruso2019}. Building on \citet{Buraimo2022}, we estimate the following linear regression model of TV demand:
\begin{equation}
\begin{split}
\ln(\text{audience size}) =\; & f\bigl(
\ln(\text{suspense}),
\text{ }\ln(\text{surprise}),
\text{ player quality}, \\
& \text{ outcome uncertainty},
\text{ championship significance}, \\
& \text{ European significance},
\text{ relegation significance},
\text{ controls}
\bigr)
\end{split}
\tag{A.1}
\end{equation}
where the control variables include home- and away-team dummies, season dummies, month dummies, a derby-match indicator, a weekday indicator, a Christmas indicator, a BT Sport indicator, and BT Sport $\times$ season interaction terms. For a detailed description of the variable construction, we refer readers to \citet{Buraimo2022}.

Table \ref{tab:tv_demand} presents the results for the TV demand model. In Column (1), we first replicate the specification of \citet{Buraimo2022} using the natural log of audience size during the match as the dependent variable, and obtain almost identical results. Most importantly for our study, the coefficient of outcome uncertainty, measured pre-match, is non-significant. In Column (2) of Table \ref{tab:tv_demand}, we add the natural log of suspense to the model. The coefficient is positive and statistically significant (\(\beta = 0.046\), \( p \) = 0.017), indicating that a 1\% increase in suspense is associated with a 0.046\% increase in audience size.

In Column (3) of Table \ref{tab:tv_demand}, we add the natural log of surprise to the baseline model. The coefficient is positive and significant (\(\beta = 0.052\), \( p \)  = 0.003), indicating that a 1\% increase in surprise is associated with a 0.052\% increase in audience size. In Column (4), we include both ln(suspense) and ln(surprise) in the model. While both coefficients remain positive, the coefficient of ln(suspense) is not estimated precisely enough to reach statistical significance at the conventional level, which may be due to the relatively low number of observations. As additional unreported robustness checks, we derive the first principal component of ln(suspense) and ln(surprise) to aggregate them into a single measure. The first principal component explains approximately 85\% of the variance contained in ln(suspense) and ln(surprise) and loads positively on both variables. The results show that the first principal component has a positive and significant effect on the audience size. Finally, we include the standardized sum of ln(suspense) and ln(surprise) in the TV demand model, and the results remain consistent.

Overall, the results of this subsample analysis suggest that both suspense and surprise are positively associated with TV demand, whereas the most commonly used measure of outcome uncertainty, based on the absolute difference in pre-match win probabilities for either team, has no effect.

\begin{table}[ht!]
    \centering
    \caption{Regression Results TV Demand}
    \label{tab:tv_demand}
    \small
    \begin{tabular}{l c c c c}
        \toprule
        & \multicolumn{4}{c}{Ln(audience)} \\
        \cmidrule(lr){2-5}
        & (1) & (2) & (3) & (4) \\
        \midrule
        Ln(suspense) && 0.046*** &  & 0.020   \\
                     && (0.017) &  & (0.020)   \\
        Ln(surprise) &&  & 0.052***   & 0.039* \\
                     &&  & (0.017)   & (0.021) \\
        Average player rating & 4.358*** & 4.546*** & 4.646*** & 4.658*** \\
                              & (1.423)  & (1.432)  & (1.432)  & (1.434)  \\
        Outcome uncertainty & $-$0.037 & $-$0.002 & $-$0.005 & 0.002 \\
                            & (0.054) & (0.054) & (0.053) & (0.054) \\
        Derby match & 0.053* & 0.049 & 0.057* & 0.054* \\
                    & (0.030) & (0.030) & (0.029) & (0.030) \\
        Match significance (championship) & 0.666*** & 0.643*** & 0.645*** & 0.640*** \\
                                          & (0.134)  & (0.135)  & (0.137)  & (0.137)  \\
        Match significance (European) & 0.193* & 0.185* & 0.196* & 0.192* \\
                                      & (0.111) & (0.111) & (0.111) & (0.111) \\
        Match significance (relegation) & 0.361*** & 0.364*** & 0.369*** & 0.368*** \\
                                        & (0.092)  & (0.093)  & (0.093)  & (0.093)  \\
        Christmas & 0.090** & 0.095*** & 0.102*** & 0.101*** \\
                  & (0.035) & (0.036) & (0.035) & (0.036) \\
        Weekday & 0.015 & 0.014 & 0.015 & 0.015 \\
                & (0.020) & (0.020) & (0.020) & (0.020) \\
        BT & $-$0.790*** & $-$0.775*** & $-$0.782*** & $-$0.778*** \\
           & (0.049)   & (0.050)   & (0.049)   & (0.050)   \\
        BT × season ending: 2015 & 0.140** & 0.127** & 0.126** & 0.124** \\
                                  & (0.060) & (0.060) & (0.059) & (0.059) \\
        BT × season ending: 2016 & 0.277*** & 0.265*** & 0.270*** & 0.266*** \\
                                  & (0.064) & (0.064) & (0.064) & (0.064) \\
        BT × season ending: 2017 & 0.299*** & 0.283*** & 0.288*** & 0.284*** \\
                                  & (0.063) & (0.064) & (0.063) & (0.064) \\
        BT × season ending: 2018 & 0.391*** & 0.387*** & 0.399*** & 0.396*** \\
                                  & (0.066) & (0.067) & (0.066) & (0.067) \\
        BT × season ending: 2019 & 0.282*** & 0.272*** & 0.273*** & 0.271*** \\
                                  & (0.066) & (0.066) & (0.065) & (0.066) \\
        \midrule
        Month FE & Yes & Yes & Yes & Yes \\
        Season FE & Yes & Yes & Yes & Yes \\
        Home team FE & Yes & Yes & Yes & Yes \\
        Away team FE & Yes & Yes & Yes & Yes \\
        Observations & 790 & 790 & 790 & 790 \\
        Adj-R\textsuperscript{2} & 0.714 & 0.716 & 0.717 & 0.717 \\
        \bottomrule
    \end{tabular}
    \caption*{\footnotesize\textit{Notes}: This table reports the results from OLS regressions with robust standard errors in parentheses. The data for all variables except suspense and surprise were provided by \citet{Buraimo2022}. *** $p < 0.01$, ** $p < 0.05$, * $p < 0.1$.}
\end{table}

\clearpage
\section{Team-Seasons Below the Benchmark}\label{sec:team_seasons}

\begin{table}[ht]
    \singlespacing
    \centering
    \caption{Premier League Team-Seasons Below Benchmark}
    \label{tab:epl_suspense_cond}
    \footnotesize
    \resizebox{\textwidth}{!}{
    \begin{tabular}{l c c c c c c c c c c c c c c c}
        \toprule
        Team & 2010 & 2011 & 2012 & 2013 & 2014 & 2015 & 2016 & 2017 & 2018 & 2019 & 2020 & 2021 & 2022 & 2023 & Count\\
\midrule
\multicolumn{16}{l}{\textit{Panel A: Suspense}} \\
        Arsenal & 5.49\tableuparrowgreen &  &  & 4.92\tableuparrowgreen & 5.42\tableuparrowgreen &  & 5.47\tableuparrowgreen & 5.28\tableuparrowgreen & 5.45\tableuparrowgreen &  &  & 4.84\tableuparrowgreen & 4.97\tableuparrowgreen & 4.97\tableuparrowgreen & 9\\
        Aston Villa &  &  &  &  &  &  &  &  &  &  &  &  &  & 5.36 & 1\\
        Bolton Wanderers &  & 5.39\tabledownarrowred &  &  &  &  &  &  &  &  &  &  &  &  & 1\\
        Bournemouth &  &  &  &  &  &  &  &  & 5.20 &  &  &  &  &  & 1\\
        Burnley &  &  &  &  &  &  &  &  & 5.45 &  &  &  &  & 5.17\tabledownarrowred & 2\\
        Cardiff City &  &  &  &  &  &  &  &  & 5.25\tabledownarrowred  &  &  &  &  &  &1 \\
        Chelsea & 5.24 &  &  &  & 5.10 &  & 4.70 & 5.30 & 5.23 & 5.14 &  &  &  &  &6 \\
        Crystal Palace &  &  &  &  &  &  &  &  &  &  & 5.39 &  &  &  & 1\\
        Fulham &  &  &  & 5.39\tabledownarrowred  &  &  &  &  & 5.17\tabledownarrowred  &  &  &  &  & 5.44& 3\\
        Huddersfield Town &  &  &  &  &  &  &  & 5.37 & 5.25\tabledownarrowred  &  &  &  &  &  & 2\\
        Hull City &  &  &  &  &  &  & 5.41\tabledownarrowred &  &  &  &  &  &  &  & 1\\
        Leicester City &  &  &  &  &  &  & 5.46 &  &  &  &  &  &  &  & 1\\
        Liverpool &  &  &  & 4.63\tableuparrowgreen &  &  &  & 4.83\tableuparrowgreen & 4.57\tableuparrowgreen & 4.61\tableuparrowgreen &  & 4.76\tableuparrowgreen &  &  & 5\\
        Manchester City &  & 5.08\tableuparrowgreen &  & 4.40\tableuparrowgreen & 5.24\tableuparrowgreen & 5.11\tableuparrowgreen & 4.92\tableuparrowgreen & 3.92\tableuparrowgreen & 3.37\tableuparrowgreen & 3.94\tableuparrowgreen & 4.39\tableuparrowgreen & 4.16\tableuparrowgreen & 4.31\tableuparrowgreen & 4.55\tableuparrowgreen & 12\\
        Manchester United &  & 4.43 & 5.38 & 5.12 &  &  &  & 5.41 & 5.33 &  &  &  &  &  & 5\\
        Newcastle United &  &  &  &  &  &  &  &  &  &  &  &  &  & 5.35 & 1\\
        Norwich City &  &  &  &  &  &  &  &  &  & 5.31\tabledownarrowred  &  & 4.99\tabledownarrowred  &  &  &2 \\
        Sheffield United &  &  &  &  &  &  &  &  &  &  &  &  &  & 4.91\tabledownarrowred  & 1\\
        Sunderland &  &  &  &  &  &  & 5.30\tabledownarrowred &  &  &  &  &  &  &  & 1\\
        Tottenham Hotspur &  &  &  &  &  &  & 5.12 & 5.05 & 5.35 &  & 5.40 & 5.29 &  &  & 5\\
        Watford &  &  &  &  &  &  &  &  & 5.48 &  &  & 5.42\tabledownarrowred  &  &  & 2\\
     \midrule
     League mean & 5.94 & 5.79 & 6.10 & 5.62 & 5.97 & 6.01 & 5.60 & 5.70 & 5.45 & 5.78 & 5.84 & 5.67 & 5.79 & 5.59 & \\ 
\midrule
\multicolumn{16}{l}{\textit{Panel B: Surprise}} \\
Arsenal &  &  &  & 1.12\tableuparrowgreen &  &  &  &  &  &  &  &  &  &  & 1\\
Huddersfield Town &  &  &  &  &  &  &  & 1.11 &  &  &  &  &  &  & 1\\
Liverpool &  &  &  &  &  &  &  &  & 1.09\tableuparrowgreen &  &  &  &  &  & 1\\
Manchester City &  &  &  &  &  &  &  & 1.03\tableuparrowgreen & 0.77\tableuparrowgreen &  & 1.11\tableuparrowgreen & 0.98\tableuparrowgreen & 1.11\tableuparrowgreen &  & 5\\
Manchester United &  & 1.06 &  &  &  &  &  &  &  &  &  &  &  &  & 1\\
Norwich City &  &  &  & 1.13\tabledownarrowred &  &  &  &  & & 1.12\tabledownarrowred  &  & 1.02\tabledownarrowred  &  &  & 3\\
Sheffield United &  &  &  &  &  &  &  &  &  &  & 1.16\tabledownarrowred  &  &  &  & 1\\
\midrule
League mean & 1.50 & 1.42 & 1.51 & 1.34 & 1.38 & 1.46 & 1.37 & 1.38 & 1.33 & 1.37 & 1.37 & 1.38 & 1.39 & 1.52 & \\
        \bottomrule
    \end{tabular}}
    \caption*{\footnotesize\textit{Notes}: Entries report team-season mean values that are significantly below the benchmark.\tableuparrowgreen indicates top teams as defined in Section 4.2;\tabledownarrowred indicates season-specific relegated teams. The league mean row reports the league-wide mean in each season. All team entries satisfy $p<0.05$.}
\end{table}

\begin{table}[ht!]
    \singlespacing
    \centering
    \caption{Bundesliga Team-Seasons Below Benchmark}
    \label{tab:bundesliga_suspense_cond}
    \footnotesize
    \resizebox{\textwidth}{!}{
    \begin{tabular}{l c c c c c c c c c c c c c c c}
        \toprule
        Team & 2010 & 2011 & 2012 & 2013 & 2014 & 2015 & 2016 & 2017 & 2018 & 2019 & 2020 & 2021 & 2022 & 2023 &Count\\
\midrule
\multicolumn{16}{l}{\textit{Panel A: Suspense}} \\
        Bayer Leverkusen &  &  &  &  &  &  &  &  & 5.27\tableuparrowgreen &  &  &  &  & 4.82\tableuparrowgreen &2\\
        Bayern Munich & 5.16\tableuparrowgreen & 4.70\tableuparrowgreen & 3.96\tableuparrowgreen & 3.72\tableuparrowgreen & 4.25\tableuparrowgreen & 4.08\tableuparrowgreen & 4.51\tableuparrowgreen & 4.00\tableuparrowgreen & 4.04\tableuparrowgreen & 3.89\tableuparrowgreen & 4.78\tableuparrowgreen & 4.38\tableuparrowgreen & 4.10\tableuparrowgreen & 4.35\tableuparrowgreen &14\\
        Bochum &  &  &  &  &  &  &  &  &  &  &  &  & 5.00 &  &1\\
        Darmstadt 98 &  &  &  &  &  &  &  &  &  &  &  &  &  & 5.05\tabledownarrowred &1\\
        Dortmund & 5.07\tableuparrowgreen & 4.98\tableuparrowgreen & 5.13\tableuparrowgreen & 5.26\tableuparrowgreen &  & 4.75\tableuparrowgreen &  &  &  & 5.14\tableuparrowgreen &  & 5.11\tableuparrowgreen & 4.88\tableuparrowgreen & & 8\\
        D\"usseldorf &  &  &  &  &  &  &  &  & 5.36 &  &  &  &  &  &1\\
        Greuther F\"urth &  &  &  &  &  &  &  &  &  &  &  & 5.23\tabledownarrowred &  &  &1\\
        Hertha BSC &  &  &  &  &  &  &  &  &  &  &  & 5.41 &  &  &1\\
        K\"oln &  & 5.31\tabledownarrowred &  &  &  &  &  &  &  & 4.94 &  &  &  &  &2\\
        Paderborn 07 &  &  &  &  &  &  &  &  &  & 5.40\tabledownarrowred &  &  &  &  &1\\
        RB Leipzig &  &  &  &  &  &  &  &  &  & 5.12 &  & 5.06 & 5.45 &  &3\\
        Schalke 04 &  & 5.13 &  &  &  &  &  &  &  &  & 4.94\tabledownarrowred &  &  &  &2\\
        Stuttgart &  &  &  &  &  &  &  &  & 5.40\tabledownarrowred &  &  &  &  & 5.28 &2\\
 \midrule       
League mean & 6.09 & 5.88 & 5.87 & 5.68 & 6.06 & 5.96 & 6.02 & 5.98 & 5.65 & 5.61 & 5.95 & 5.69 & 5.62 & 5.80 &\\
\midrule
\multicolumn{16}{l}{\textit{Panel B: Surprise}} \\
        Arminia &  &  &  &  &  &  &  &  &  &  & 1.11 &  &  &  &1\\
        Augsburg &  &  & 1.13 &  &  &  &  &  &  &  &  &  &  &  &1\\
        Bayern Munich &  & 1.00\tableuparrowgreen & 1.00\tableuparrowgreen & 0.98\tableuparrowgreen & 0.91\tableuparrowgreen & 0.93\tableuparrowgreen &  & 1.07\tableuparrowgreen & 1.10\tableuparrowgreen & 1.06\tableuparrowgreen &  & 1.13\tableuparrowgreen & 1.11\tableuparrowgreen &  &10\\
 \midrule           
League mean & 1.54 & 1.41 & 1.43 & 1.43 & 1.43 & 1.43 & 1.47 & 1.41 & 1.42 & 1.40 & 1.45 & 1.44 & 1.40 & 1.47 &\\
        \bottomrule
    \end{tabular}}
    \caption*{\footnotesize\textit{Notes}: Entries report team-season mean values that are significantly below the benchmark.\tableuparrowgreen indicates top teams as defined in Section 4.3;\tabledownarrowred indicates season-specific relegated teams. The league mean row reports the league-wide mean in each season. All team entries satisfy $p<0.05$.}
\end{table}

\begin{table}[ht!]
    \singlespacing
    \centering
    \caption{La Liga Team-Seasons Below Benchmark}
    \label{tab:laliga_suspense_cond}
    \footnotesize
    \resizebox{\textwidth}{!}{
    \begin{tabular}{l c c c c c c c c c c c c c c c}
        \toprule
        Team & 2010 & 2011 & 2012 & 2013 & 2014 & 2015 & 2016 & 2017 & 2018 & 2019 & 2020 & 2021 & 2022 & 2023 &Count\\
\midrule
\multicolumn{16}{l}{\textit{Panel A: Suspense}} \\
        Alavés &  &  &  &  &  &  &  & 5.36 &  &  &  &  &  &  &1\\
        Athletic Club & 5.51 &  &  &  &  & 5.22 &  &  &  &  &  &  &  & 5.42 &3\\
        Atlético Madrid & 5.24\tableuparrowgreen &  &  & 4.87\tableuparrowgreen & 4.81\tableuparrowgreen & 5.47\tableuparrowgreen & 5.34\tableuparrowgreen &  &  &  &  &  &  & 5.39\tableuparrowgreen &6\\
        Barcelona & 3.98\tableuparrowgreen & 3.64\tableuparrowgreen & 3.96\tableuparrowgreen & 4.32\tableuparrowgreen & 3.88\tableuparrowgreen & 3.58\tableuparrowgreen & 4.18\tableuparrowgreen & 4.35\tableuparrowgreen & 5.16\tableuparrowgreen & 4.72\tableuparrowgreen & 5.22\tableuparrowgreen &  & 4.74\tableuparrowgreen & & 12\\
        Eibar &  &  &  &  &  & 5.40 &  &  &  &  &  &  &  &  &1\\
        Elche &  &  &  &  & 5.31\tabledownarrowred &  &  &  &  &  &  &  &  &  &1\\
        Getafe & 5.20 &  &  &  &  & 5.22\tabledownarrowred &  &  &  &  &  &  &  &  &2\\
        Granada &  &  &  &  &  & 5.46 & 5.27\tabledownarrowred &  &  &  &  &  &  & 5.33\tabledownarrowred &3\\
        Las Palmas &  &  &  &  &  &  & 5.37 &  &  &  &  &  &  &  &1\\
        Levante &  &  &  &  & 5.32 &  &  &  &  &  &  &  &  &  &1\\
        Mallorca &  &  &  &  &  &  &  &  &  & 5.22\tabledownarrowred &  &  &  &  &1\\
        Málaga & 5.21 &  &  &  &  &  &  &  &  &  &  &  &  &  &1\\
        Osasuna &  &  &  &  &  &  & 5.16\tabledownarrowred &  &  &  &  &  &  &  &1\\
        Rayo Vallecano &  & 5.29 &  & 4.71 &  &  &  &  &  &  &  &  &  &  &2\\
        Real Madrid & 4.13\tableuparrowgreen & 3.69\tableuparrowgreen & 4.88\tableuparrowgreen & 4.29\tableuparrowgreen & 3.48\tableuparrowgreen & 4.04\tableuparrowgreen & 4.34\tableuparrowgreen & 4.68\tableuparrowgreen & 5.52\tableuparrowgreen &  &  &  &  & 5.26\tableuparrowgreen &10\\
        Real Sociedad &  &  &  &  &  &  & 5.52 &  &  &  &  &  &  &  &1\\
        Sevilla &  &  &  &  & 5.20 &  &  &  &  &  &  &  &  &  &1\\
        \midrule
League mean & 5.57 & 5.73 & 5.90 & 5.66 & 5.56 & 5.66 & 5.65 & 5.77 & 6.22 & 6.13 & 6.17 & 6.20 & 6.08 & 6.16 &\\
\midrule
\multicolumn{16}{l}{\textit{Panel B: Surprise}} \\
        Alavés &  &  &  &  &  &  &  & 1.14 &  &  &  &  &  & & 1\\
        Athletic Club &  &  &  &  &  & 1.15 &  &  &  &  &  &  &  &  &1\\
        Atlético Madrid &  &  &  & 1.07\tableuparrowgreen & 1.08\tableuparrowgreen &  &  &  &  &  &  &  &  &  &2\\
        Barcelona & 0.90\tableuparrowgreen & 1.04\tableuparrowgreen &  &  & 0.90\tableuparrowgreen & 0.88\tableuparrowgreen &  &  &  &  &  &  &  & 0.95\tableuparrowgreen &5\\
        Getafe &  &  &  &  &  &  &  &  &  &  & 1.19 &  &  &  &1\\
        Osasuna &  &  &  & 1.09\tabledownarrowred &  &  &  &  &  &  &  &  &  &  &1\\
        Real Madrid & 0.96\tableuparrowgreen & 1.02\tableuparrowgreen &  &  & 0.99\tableuparrowgreen & 1.03\tableuparrowgreen &  &  &  &  &  &  &  &  &4\\
        \midrule
League mean & 1.31 & 1.38 & 1.42 & 1.32 & 1.30 & 1.32 & 1.44 & 1.34 & 1.51 & 1.44 & 1.47 & 1.49 & 1.43 & 1.53 &\\
        \bottomrule
    \end{tabular}}
    \caption*{\footnotesize\textit{Notes}: Entries report team-season mean values that are significantly below the benchmark.\tableuparrowgreen indicates top teams as defined in Section 4.4;\tabledownarrowred indicates season-specific relegated teams. The league mean row reports the league-wide mean in each season. All team entries satisfy $p<0.05$.}
\end{table}

\begin{table}[ht!]
    \singlespacing
    \centering
    \caption{Serie A Team-Seasons Below Benchmark}
    \label{tab:seriea_suspense_cond}
    \footnotesize
    \resizebox{\textwidth}{!}{
    \begin{tabular}{l c c c c c c c c c c c c c c c}
        \toprule
        Team & 2010 & 2011 & 2012 & 2013 & 2014 & 2015 & 2016 & 2017 & 2018 & 2019 & 2020 & 2021 & 2022 & 2023 & Count\\
\midrule
\multicolumn{16}{l}{\textit{Panel A: Suspense}} \\
        Atalanta &  &  &  &  &  &  &  &  &  &  & 4.74 &  &  &  &1\\
        Benevento &  &  &  &  &  &  &  & 5.17\tabledownarrowred &  &  &  &  &  &  &1\\
        Bologna &  &  &  &  &  &  &  &  &  &  & 5.46 &  &  &  &1\\
        Cagliari &  &  &  &  &  &  & 4.86 &  &  &  &  &  &  &  &1\\
        Crotone &  &  &  &  &  &  &  & 5.08\tabledownarrowred &  &  & 5.06\tabledownarrowred &  &  &  &2\\
        Fiorentina &  &  & 5.48 &  &  &  &  &  &  &  &  &  &  &  &1\\
        Frosinone &  &  &   &  &  &  &  &  & 5.28\tabledownarrowred &  &  &  &  &  &1\\
        Hellas Verona &  &  &  & 5.39 &  &  &  & 4.67\tabledownarrowred &  &  &  &  &  &  &2\\
        Internazionale &  &  &  &  &  &  &  &  &  & 5.36\tableuparrowgreen & 5.24\tableuparrowgreen & 5.08\tableuparrowgreen &  & 4.89\tableuparrowgreen &4\\
        Juventus &  &  & 5.44\tableuparrowgreen & 4.45\tableuparrowgreen & 5.39\tableuparrowgreen & 4.91\tableuparrowgreen & 4.43\tableuparrowgreen & 4.76\tableuparrowgreen & 5.15\tableuparrowgreen & 5.29\tableuparrowgreen & 5.24\tableuparrowgreen &  &  &  &9\\
        Lazio &  &  & 5.30 &  &  & 5.49 & 5.41 &  &  &  & 5.44 & 5.22 &  &  &5\\
        Lecce &  &  &  &  &  &  &  &  &  & 5.37\tabledownarrowred &  &  &  &  &1\\
        Livorno &  &  &  & 5.09\tabledownarrowred &  &  &  &  &  &  &  &  &  &  &1\\
        Milan &  & 5.28 &  &  &  &  &  &  &  &  &  & 5.33 &  &  &2\\
        Napoli &  &  &  & 5.18\tableuparrowgreen &  & 4.97\tableuparrowgreen & 4.52\tableuparrowgreen & 4.92\tableuparrowgreen & 5.40\tableuparrowgreen &  & 4.88\tableuparrowgreen &  & 5.19\tableuparrowgreen &  &7\\
        Pescara &  &  & 4.79\tabledownarrowred &  &  &  & 5.14\tabledownarrowred &  &  &  &  &  &  & & 2\\
        Roma &  &  &  & 5.11 &  &  & 4.91 &  &  &  & 5.40 &  &  & & 3\\
        Salernitana &  &  &  &  &  &  &  &  &  &  &  & 5.33 &  & & 1\\
        Sampdoria &  &  &  & 5.44 &  & 5.31 &  &  &  &  &  & 5.05 &  &  &3\\
        Udinese &  &  &  &  &  &  &  & 5.48 &  &  &  &  &  &  &1\\
        \midrule
        League mean & 6.19 & 5.96 & 6.01 & 5.77 & 6.11 & 5.95 & 5.60 & 5.68 & 5.97 & 5.87 & 5.71 & 5.80 & 6.10 & 6.12 &\\
\midrule
\multicolumn{16}{l}{\textit{Panel B: Surprise}} \\
        Atalanta &  &  &  &  &  &  &  &  &  &  & 1.16 &  &  &  &1\\
        Cagliari &  & 1.07 &  &  &  &  &  &  &  &  &  &  &  &  &1\\
        Hellas Verona &  &  &  &  &  &  &  & 1.12\tabledownarrowred &  &  &  &  &  &  &1\\
        Internazionale &  &  &  &  &  &  &  &  &  &  &  &  &  & 1.12\tableuparrowgreen &1\\
        Juventus &  & 1.17\tableuparrowgreen &  & 1.07\tableuparrowgreen &  & 1.10\tableuparrowgreen &  &  &  &  &  &  &  &  &3\\
        Napoli &  &  &  &  &  & 1.11\tableuparrowgreen &  &  &  &  &  &  &  &  &1\\
        Pescara &  &  & 1.09\tabledownarrowred &  &  &  &  &  &  &  &  &  &  &  &1\\
        Roma &  &  &  & 1.12 &  &  &  & 1.06 &  &  &  &  &  &  &2\\
        \midrule
        League mean & 1.45 & 1.36 & 1.44 & 1.43 & 1.48 & 1.36 & 1.41 & 1.34 & 1.44 & 1.50 & 1.45 & 1.46 & 1.44 & 1.49 &\\
        \bottomrule
    \end{tabular}}
    \caption*{\footnotesize\textit{Notes}: Entries report team-season mean values that are significantly below the benchmark.\tableuparrowgreen indicates top teams as defined in Section 4.5;\tabledownarrowred indicates season-specific relegated teams. The league mean row reports the league-wide mean in each season. All team entries satisfy $p<0.05$.}
\end{table}

\begin{table}[ht!]
    \singlespacing
    \centering
    \caption{Ligue 1 Team-Seasons Below Benchmark}
    \label{tab:ligue1_suspense_cond}
    \footnotesize
    \resizebox{\textwidth}{!}{
    \begin{tabular}{l c c c c c c c c c c c c c c c}
        \toprule
        Team & 2010 & 2011 & 2012 & 2013 & 2014 & 2015 & 2016 & 2017 & 2018 & 2019 & 2020 & 2021 & 2022 & 2023 & Count\\
\midrule
\multicolumn{16}{l}{\textit{Panel A: Suspense}} \\
        Ajaccio &  &  &  &  &  &  &  &  &  &  &  &  & 5.31\tabledownarrowred &  &1\\
        Angers &  &  &  &  &  &  &  &  &  &  &  &  & 5.33\tabledownarrowred &  &1\\
        Dijon &  &  &  &  &  &  &  &  &  &  & 5.42\tabledownarrowred &  &  &  &1\\
        Lyon &  &  &  &  &  &  &  & 5.35\tableuparrowgreen &  &  & 5.34\tableuparrowgreen &  &  & & 2\\
        Marseille &  &  &  &  &  &  &  & 5.27\tableuparrowgreen &  &  &  &  &  &  &1\\
        Monaco &  &  &  &  &  &  & 4.38 & 5.12 &  &  &  &  &  &  &2\\
        Paris Saint-Germain &  &  &  & 4.80\tableuparrowgreen & 4.95\tableuparrowgreen & 4.16\tableuparrowgreen & 4.31\tableuparrowgreen & 3.82\tableuparrowgreen & 4.14\tableuparrowgreen & 3.88\tableuparrowgreen & 4.46\tableuparrowgreen & 4.62\tableuparrowgreen & 4.53\tableuparrowgreen & 5.07\tableuparrowgreen &11\\
        Rennes &  &  &  &  &  &  &  &  &  &  &  & 5.23 &  &  &1\\
        \midrule
        League mean & 6.35 & 6.22 & 6.15 & 6.01 & 6.04 & 6.03 & 5.86 & 5.87 & 6.17 & 6.09 & 5.89 & 6.10 & 5.83 & 6.08 &\\
\midrule
\multicolumn{16}{l}{\textit{Panel B: Surprise}} \\
        Arles-Avignon & 1.15\tabledownarrowred &  &  &  &  &  &  &  &  &  &  &  &  &  &1\\
        Dijon &  &  &  &  &  &  &  &  &  &  & 1.12\tabledownarrowred &  &  &  &1\\
        Nancy &  &  &  &  &  &  & 1.13\tabledownarrowred &  &  &  &  &  &  &  &1\\
        Nice &  &  &  & 1.14 &  &  &  &  & 1.17 &  &  &  &  &  &2\\
        Paris Saint-Germain &  &  &  &  &  & 0.96\tableuparrowgreen & 1.12\tableuparrowgreen & 1.01\tableuparrowgreen &  &  & 1.03\tableuparrowgreen &  &  & & 4\\
        \midrule
        League mean & 1.44 & 1.50 & 1.45 & 1.38 & 1.41 & 1.39 & 1.35 & 1.42 & 1.44 & 1.41 & 1.40 & 1.49 & 1.40 & 1.45 &\\
        \bottomrule
    \end{tabular}}
    \caption*{\footnotesize\textit{Notes}: Entries report team-season mean values that are significantly below the benchmark.\tableuparrowgreen indicates top teams as defined in Section 4.6;\tabledownarrowred indicates season-specific relegated teams. The league mean row reports the league-wide mean in each season. All team entries satisfy $p<0.05$.}
\end{table}

\end{document}